\documentclass[manuscript]{aastex631}

\usepackage{newtxtext,newtxmath}
\usepackage[T1]{fontenc}
\usepackage{graphicx}
\usepackage{lipsum}
\usepackage{hyperref}
\usepackage{longtable}
\usepackage[figuresright]{rotating}
\usepackage{multirow}

\begin{document}

\title{Density Profiles of \texttt{TNG300} Voids across Cosmic Time}

\correspondingauthor{Olivia Curtis}
\email{ocurtis@psu.edu}

\author[0000-0002-0212-4563]{Olivia Curtis}
\affiliation{Department of Astronomy \& Astrophysics, The Pennsylvania State University, 251 Pollock Road, University Park, PA 16802, USA}\affiliation{Department of Astronomy \& Institute for Astrophysical Research, 725 Commonwealth Ave., Boston University, Boston, MA 02215, USA}

\author[0000-0001-6928-4345]{Bryanne McDonough}
\affiliation{Department of Physics, Northeastern University, 360 Huntington Ave, Boston, MA 02115, USA}\affiliation{Department of Astronomy \& Institute for Astrophysical Research, 725 Commonwealth Ave., Boston University, Boston, MA 02215, USA}

\author[0000-0001-7917-7623]{Tereasa G. Brainerd}
\affiliation{Department of Astronomy \& Institute for Astrophysical Research, 725 Commonwealth Ave., Boston University, Boston, MA 02215, USA}

\begin{abstract}

We present radial density profiles, as traced by luminous galaxies and dark matter particles, for voids in eleven snapshots of the \texttt{TNG300} simulation.  The snapshots span 11.65~Gyr of cosmic time, corresponding to the redshift range $0 \le z \le 3$. Using the comoving galaxy fields, voids were identified via a well-tested, watershed transformation-based algorithm. Voids were defined to be underdense regions that are unlikely to have arisen from Poisson noise, resulting in the selection of $\sim100-200$ of the largest underdense regions in each snapshot. At all redshifts, the radial density profiles as traced by both the galaxies and the dark matter resemble inverse top-hat functions. However, details of the functions (particularly the underdensities of the innermost regions and the overdensities of the ridges) evolve considerably more for the dark matter density profiles than for the galaxy density profiles. At all redshifts, a linear relationship between the galaxy and dark matter density profiles exists, and the slope of the relationship is similar to the bias estimates for \texttt{TNG300} snapshots. Lastly, we identify distinct environments in which voids can exist, defining ``void-in-void" and ``void-in-cloud" populations (i.e., voids that reside in larger underdense or overdense regions, respectively) and we investigate ways in which the relative densities of dark matter and galaxies in the interiors and ridges of these structures vary as a function of void environment.
%
%
%
%

\end{abstract}

\keywords{Large-scale structure of the Universe (902) -- Voids (1779) -- Magnetohydrodynamical simulations (1966)}


\section{Introduction}
\label{sec:intro}

Cosmic voids are the largest and emptiest structures in the universe (\citealt{tifft1976}; \citealt{gregory1978}). 
They are also excellent cosmological probes due to the fact that, regardless of how a tracer population is defined, the positions and motions of tracers within voids adhere to linear theory at all size scales (e.g., \citealt{colberg2005}; 
\citealt{nadathur2015}; \citealt{schuster2023}). Because of this, voids have been used for a number of cosmological purposes in both observations and numerical simulations, including estimations of the sum of the neutrino masses (e.g., \citealt{villaescusanavarro2013}; 
\citealt{schuster2019};  
\citealt{kreisch2022}), tests of modified theories of gravity (e.g., \citealt{li2012}; \citealt{clampitt2013}; 
\citealt{zivick2015}; 
\citealt{contarini2021}), constraints on the dark energy equation of state (e.g., \citealt{lee2009}; \citealt{lavaux2010};
\citealt{adermann2017}), deducing values of cosmological parameters (e.g., \citealt{park2007}; \citealt{hamaus2015};
\citealt{pelliciari2023}; \citealt{wang2024}), and characterizing the impact of voids on the integrated Sachs-Wolfe effect for the cosmic microwave background (e.g., \citealt{kovac2014}; \citealt{nadathur2014};
\citealt{cai2017};
\citealt{owusu2023}). 


Voids are interconnected with the flow patterns of the cosmic web (e.g., \citealt{bothun1992}; \citealt{tully2019}; \citealt{vallesperez2021}; \citealt{courtois2023}; \citealt{schuster2023}). \cite{courtois2023} used Cosmicflows-3 \citep{tully2016} data to study void galaxies in the Calar Alto Void Integral-field Treasury surveY (see, e.g., \citealt{cavity1}; \citealt{cavity2}) and \cite{tully2019} used Cosmicflows-3 to study the dynamics of the Local Void \citep{Tully1987}, from which they determined that the repulsion from the Local Void greatly contributes to the motion of the Local Group. Both of these studies showed that, in the local universe, the flow in and around voids is primarily dominated by the dynamical tidal influence of nearby filaments and walls, sometimes resulting in a flow field that enters the void. Therefore, an understanding of the relative densities of dark matter and galaxies in the ridges of voids could help model these flow patterns in the future.

Void density profiles resemble reverse spherical top-hats (\citealt{sheth}; \citealt{hamaus2014}; \citealt{Ricciardelli2014}); that is, they have very underdense centers and ridges that are $\sim20\%$ more overdense than the average of the survey from which they were obtained. In both simulations and observations, void density profiles have been shown to exhibit this shape regardless of tracer type, and their averaged radial and integrated number density profiles are well modeled by empirical formulae (e.g., \citealt{paz2013}; \citealt{hamaus2014}; \citealt{pollina2017};  \citealt{schuster2023}). 
Understanding the mass density fields of voids is especially important for studies of redshift space distortions around voids (e.g., \citealt{hamaus2015}; \citealt{correa2021}; \citealt{hamaus2022}). 

Furthermore, \cite{kaiser1984} showed that galaxy distributions do not directly mirror the dark matter distribution. That is, on the largest scales, there is a shift between the clustering amplitudes of void tracers and the underlying mass field, where this shift is known as a linear bias when it takes on a constant value. 
\cite{pollina2017} found that there is a linear relationship between the dark matter field and the void tracer field from a study of voids in the $z=0.14~$ \texttt{Magneticum Pathfinder} simulation (see, e.g., \citealt{dolag2016} and \citealt{hirschmann2014}) and that the slope of the relationship is a good estimate for the clustering bias. Observational confirmation of the relationship found by \cite{pollina2017} may be possible in the future; this will, however, depend upon precise measurements of weak lensing by voids, allowing direct constraints on their mass distributions to be obtained (see, e.g., \citealt{izumi2013}; \citealt{barreira2015}; \citealt{sanchez2016cosmic}; \citealt{paillas2019}; \citealt{bonici2023}). 

{Voids also play an important role in modulating cosmic flow patterns (see, e.g., \citealt{cautun2014}). In the simplest picture, voids expand and expel material outward into sheets that then drive the material toward filaments and clusters (\citealt{zeldovich1970}; \citealt{shandarin1989}; \citealt{vanhaarlem1993}). Several studies have analyzed the evolution voids and how these structures interact with the cosmic web as a whole. For example, \cite{cautun2014} tracked the mass distributions, velocity fields, and structure sizes within the \texttt{Millennium} simulation \citep{springel2005}, showing how the mass and volume of the cosmic web evolved from being dominated by tenuous sheets and filaments at early times to being dominated in volume by voids and in mass by clusters and filaments at later times. They also showed that, compared to halos and filaments in more dense environments at low redshifts, void halos are less massive and void filaments are wider. They attribute this ``frozen'' structure formation to the fact that void filaments and halos are too far apart and too feeble in mass to merge with neighboring structures, halting the development of medium-scale density fluctuations at low redshifts \citep{aragoncalvo2013}.}

{Other authors have tracked the evolution of individual voids across cosmic time (see, e.g., \citealt{sutter2014c}; \citealt{wojtak2016}). \cite{sutter2014c} achieved this by tracking the IDs of dark matter particles and identifying which voids share particles from snapshot to snapshot. Using this method, \cite{sutter2014c} found that typical voids form early, rarely merge, and grow slowly. They also investigated void evolution in terms of the void hierarchy (e.g., \citealt{sheth}), tracking how voids embedded in overdense regions can collapse over time and showing that, compared to voids in denser environments, voids in underdense regions experience more mergers. However, \cite{wojtak2016} argued that this method of constructing merger trees introduces ambiguity as most particles tend to lie on the boundary of a void and therefore can be exchanged with neighboring voids, leading to spurious reconfigurations of the shapes and positions of voids. Instead, \cite{wojtak2016} constructed a single void catalog at $z=0.0$ using a watershed void finding algorithm \citep{platen2007} and tracked each watershed basin backward in time to identify the corresponding voids at earlier redshifts. They showed that while the axis ratios of individual voids can change over time by a factor of $\sim0.2$, the overall distribution of void shapes remains the same due to the fact that the more elongated voids tend to become more spherical and vice versa. \cite{wojtak2016} also showed that the orientations of the void principal axes remain constant over time and that the major and minor axes of nearby voids tend to align with one another at all redshifts.}

{Lastly, while the depletion of gas from voids has been seen in simulations (\citealt{vandeweygaert1993}; \citealt{padilla2005}) and in observations (see, e.g., \citealt{ceccarelli2006}; \citealt{patiri2012}; \citealt{tully2019}), \cite{vallesperez2021} showed that at $z=0$, around $10\%$ of the mass within voids may be accreted from surrounding overdense regions. By analyzing the velocity fields around voids from $z=100$ to $z=0$, they show that it is possible for larger surrounding clouds of material to collapse into voids 
and for sheets and filaments to pass through the outer regions of voids as they collapse towards dense environments. They also argue that it is possible for events such as galaxy mergers (e.g.,   \citealt{behroozi2013}) and strong shocks (e.g., \citealt{zhang2020}) to unbind material from massive halos and send it streaming into voids. These results disrupt the idea that voids are pristine environments and are counterintuitive to the picture that voids are simply expanding and expelling material.}



In this paper, we present radial profiles of simulated voids that span the redshift range $0.0 \le z \le 3.0$. We use the \texttt{TNG300} $\Lambda$ Cold Dark Matter simulation to determine the radial dependencies of the galaxy number densities and the dark matter mass densities inside the voids, with the goal of measuring the degree to which galaxies trace the underlying mass fields. \texttt{TNG300} is the largest simulation box in the IllustrisTNG suite of simulations \citep{illustris1,illustris2,illustris3,illustris4,illustris5} and will be discussed further below. We also investigate the effect of void hierarchy on the relative distributions of dark matter and galaxies within the voids. Lastly, we extend the work done by \cite{pollina2017} to $z=3.0$ by investigating the relationships between void density profiles, as traced separately by the galaxies and the dark matter, and we compare the slopes of these relationships to clustering bias estimates in \texttt{TNG300} that were obtained by \cite{illustris4}. 

Recently, \cite{rodriguesmedrano2024} released \texttt{TNG300} void catalogs that were obtained with the 3D Spherical Void Finder (3DSVF) developed by \cite{paz2023}. Rather than a 3DSVF, here we elect to use the watershed voidfinding algorithm known as ZOBOV (ZOnes Bordering On Voidness; \citealt{neyrinck2008}). We make this choice because a study by \cite{douglass2023} showed that, compared to the radial profiles of ZOBOV-identified voids, 3DSVFs produce radial density profiles with [1] flatter interiors and [2] ridges with densities equal to the mean density of the survey in which the voids were found. Since ZOBOV makes no assumptions about void shapes, and also includes the shell-crossing regime in its void profiles, this makes ZOBOV void catalogs well suited for studies of void shapes and dynamics, whereas 3DSVFs are a better option for the study of void galaxies. 

The paper is organized as follows. \texttt{TNG300} and the snapshots that we analyze are discussed in \S\ref{sec:TNG300snaps}. We describe our void-finding algorithm in \S\ref{sec:voidfinder} and the void catalogs in \S\ref{sec:voidcats}. In \S\ref{sec:voidradprofs:radial}, we construct radial number density profiles for the voids using both the dark matter particles and the galaxies as density tracers, and we discuss how the profiles evolve over time. We then divide the voids into two populations based on their hierarchy in \S\ref{sec:voidradprofs:integrated} and we investigate the dependence of the density profiles on void hierarchy.  We summarize and discuss our results in \S\ref{sec:discussion}. 

\section{\texttt{TNG300} Snapshots}
\label{sec:TNG300snaps}

We search for voids in {the comoving galaxy fields} of eleven snapshots of the highest-resolution \texttt{TNG300} simulation. 
\texttt{TNG300} encompasses a co-moving volume of $205^3 h^{-3} \rm{Mpc}^3$. {The simulation starts at $z=127$} with $2,500^3$ dark matter particles of mass $m_{\rm dm} = 4.0 \times 10^7 h^{-1} M_\odot$ and $2,500^3$ hydrodynamical gas cells with initial baryonic mass $m_{\rm b} = 7.5 \times 10^6 h^{-1} M_\odot$ were used. 
The redshifts, snapshot numbers, lookback times, and softening lengths for each snapshot that we use are summarized in Columns 1-4 of Table~\ref{tab:voidstats}. The cosmological parameters adopted in the simulation are: $\Omega_{\Lambda,0} = 0.6911$,
$\Omega_{m,0} = 0.3089$, $\Omega_{b,0} = 0.0486$, $\sigma_8 = 0.8159$, $n_s = 0.9667$, and
$h = 0.6774$ (e.g., \citealt{Planck15}). We adopt the subhalo magnitudes from the supplementary catalog of \citet{illustris1} to assign luminosities to the galaxies. Compared to the main \texttt{TNG} subhalo catalog, the \citet{illustris1} catalog better resembles SDSS photometry because it includes the effects of dust obscuration on the simulated galaxies. 

We use the publicly available \texttt{TNG300} subhalo catalog
to identify galaxies. We {only} search for voids in the comoving galaxy field of each snapshot. 
We elect to use as a tracer population all subhalos that meet a magnitude and mass cut described below. The sparsity and bias of tracers influence void detection such that voids that are identified in a sparse and heavily biased tracer population will yield fewer voids, which are larger on average, compared to those produced using a less biased tracer population (\citealt{sutter2014b}; \citealt{nadathur2015c}; \citealt{pollina2016}; \citealt{pollina2017}). The relatively small box-size of \texttt{TNG300} is the limiting factor for our study as the most biased tracers in the simulation are too rare to properly trace the large-scale structure. Indeed, at $z=0.0$, there are only $\sim240$ groups more massive than $10^{14}h^{-1}M_\odot$ (i.e., objects the \texttt{TNG} team considers to be galaxy clusters). The small volume of \texttt{TNG300} also limits the total number of voids that we can identify. An advantage of this simulation is that it generates subhalos which precisely reproduce global galaxy clustering \citep{illustris4} and population statistics \citep{illustris1}, in part due to its precise mass resolution elements. We thus expect to find fewer voids that are smaller compared to studies that have identified voids in other hydrodynamical simulations (see, e.g., \citealt{schuster2023}), but we will be able to better predict the distribution of galaxies within and around voids.

{A robust description of baryonic processes in hydrodynamic simulations requires resolution elements that are capable of modeling the small-scale radiative and feedback processes that affect galaxy evolution (\citealt{illustris1}). Void catalogs have historically been made under the assumption that galaxy voids can be reliably compared to void catalogs built on halos in dark matter simulations (\citealt{panchal2020}), but modern simulations have allowed for resolved subhalo void catalogs to be constructed from boxes $\gtrsim 100$Mpc, enabling the effects that baryonic processes and tracer selection have on void statistics to be measured directly. (see, e.g., \citealt{paillas2017}; \citealt{habouzit}; \citealt{panchal2020}; \citealt{rodriguez2022}; \citealt{schuster2023}; \citealt{curtis2024}; \citealt{rodriguesmedrano2024}).} {To construct our tracer population, we first} exclude all luminous objects that are flagged as non-cosmological in origin {(i.e., those with \texttt{SubhaloFlag} set to 0). These are objects identified by the \texttt{Subfind} algorithm \citep{springel2001} that may not have formed due to large-scale structure formation or collapse and are instead fragments of material that have formed within an existing halo through baryonic processes and are usually low mass, baryon dominated, and located near the centers of their host halos.\footnote{For more details, see \url{https://www.tng-project.org/data/docs/background/\#subhaloflag}}} With the exception of the $z=3.0$ snapshot, we also exclude galaxies with absolute $r$-band luminosities $M_r \geq -14.5$ and stellar masses $\leq 10^8h^{-1}M_\odot$. {Here, stellar mass is defined as the total mass of the stellar particles that are bound to a subhalo. We chose to use a cut on stellar mass rather than a cut on total mass as the former is an observable quantity that can be obtained by modeling galaxy spectral energy distributions.} Our chosen cut ensures that each galaxy contains a reasonable minimum number of stellar particles ($\sim 15$ for $z\leq 2.0$) and that there are enough galaxies present in the snapshot for the void finding algorithm to converge on a solution. At $z=3.0$, relatively few galaxies exist in \texttt{TNG300}; therefore, in order to have a sufficient number of galaxies such that we can identify voids at $z=3.0$, we adopt a minimum stellar mass of $10^{7.75}h^{-1}M_\odot$, corresponding to $\sim 10$ stellar particles, for this snapshot. 

{Void shapes and sizes are known to depend on tracer density (see \citealt{sutter2014b}). As our tracer number densities decrease from $\sim0.040h^3\rm{Mpc^{-3}}$ at $z=0.0$ to $\sim0.037h^3\rm{Mpc^{-3}}$ at $z=0.7$, $\sim 0.027h^3\rm{Mpc^{-3}}$ at $z=1.5$, and $0.014h^3\rm{Mpc^{-3}}$ at $z=3.0$, we expect to find larger voids at earlier redshifts. As \cite{sutter2014b} note, at $z=0$ when dark matter and halos are used as tracers, sparser sampling primarily produces voids with slightly denser ridges. These results were more pronounced for voids larger than $30h^{-1}$ Mpc and for tracer densities ($n_t$) sparser than $n_{t}\lesssim10^{-3}$. Compared to our catalogs at redshifts $< 1$, we might expect our catalogs at redshifts $>1$ to have denser ridges. However, given the fact that we have relatively high tracer densities that produce galaxy voids with radii $\sim 15-20h^{-1}$Mpc, we do not expect these mass cuts to have a significant effect on void radial profiles. Moreover, studies that identify voids out to high redshifts (see, e.g., \citealt{sutter2012}; \citealt{nadathur2016}; \citealt{mao2017}) must overcome a similar falloff in tracer density with redshift due to Malmquist bias. We thus expect our catalogs, when viewed from $z=3.0$ to $z=0.0$, to better reflect such attempts at void identification.}



\begin{sidewaystable}[!htbp]
    \centering
    \caption{TNG300 Snapshots and Void Statistics}
    \label{tab:voidstats}
    \begin{tabular}{c c c c c c c c c c c}
    \hline

    Redshift & Snap. No. & $t_{\rm lookback}$ (Gyr) & $r_{\rm softening}$ ($h^{-1}$kpc) & $N_{\rm gal}$ & $n_t$ $(h^3\rm{Mpc^{-3}})$& $N_{\rm voids}$ & $N_{\rm voids,significant}$ & Volume & $N_{\rm VIC}$ & $N_{\rm VIV}$ \\

    0.0 & 99 & 0.00  & 1.0 & 342635 & $0.040$ & 1065 & 174 & 39.5$\%$ & 88 & 86 \\
    0.1 & 91 & 1.34  & 1.1 & 338602 & $0.039$ & 1056 & 178 & 37.1$\%$ & 90 & 88 \\
    0.2 & 84 & 2.51  & 1.2 & 334219 & $0.039$ & 1075 & 166 & 35.4$\%$ & 92 & 74 \\
    0.3 & 78 & 3.53  & 1.3 & 329227 & $0.038$ & 1047 & 159 & 33.7$\%$ & 88 & 71 \\
    0.4 & 72 & 4.41  & 1.4 & 325406 & $0.038$ & 1059 & 158 & 33.0$\%$ & 91 & 67 \\
    0.5 & 67 & 5.19  & 1.5 & 322657 & $0.037$ & 1091 & 145 & 30.4$\%$ & 85 & 60 \\
    0.7 & 59 & 6.49  & 1.7 & 315329 & $0.037$ & 1093 & 165 & 32.9$\%$ & 96 & 69 \\
    1.0 & 50 & 7.94  & 2.0 & 295677 & $0.034$ & 1059 & 173 & 40.0$\%$ & 106 & 67 \\
    1.5 & 40 & 9.52  & 2.0 & 229435 & $0.027$ & 935  & 123 & 31.6$\%$ & 63 & 60 \\
    2.0 & 33 & 10.51 & 2.0 & 170014 & $0.020$ & 750  & 96  & 28.0$\%$ & 52 & 44 \\
    3.0 & 25 & 11.65 & 2.0 & 118841 & $0.014$ & 607  & 70  & 29.6$\%$ & 36 & 34 \\
    
    \hline
    \end{tabular}
\end{sidewaystable}

    %

\section{Voidfinding Algorithm}
\label{sec:voidfinder}

We use the ZOnes Bordering On Voidness (ZOBOV) algorithm (\citealt{neyrinck2008}) to indentify \texttt{TNG300} voids. ZOBOV is based on the Watershed Void Finder \citep{platen2007} and finds voids using a Voronoi Tesselation Field Estimator \citep{schaap2007}. The algorithm makes no assumptions about the shape or symmetry of voids, which allows for amorphously shaped voids to be identified. Throughout, we implement ZOBOV using the publicly available code \texttt{REVOLVER}\footnote{https://github.com/seshnadathur/Revolver} \citep{nadathur2019}, and we summarize the algorithm below.


First, the code determines the local density around each galaxy using a Voronoi Tessellation Field Estimation (see, e.g., \citealt{schaap2007}). The Voronoi tessellation draws cells around each galaxy so that each cell contains one galaxy and the regions of space that are closest to this galaxy. The density estimate around galaxy $i$ is then $1/V(i)$ where $V(i)$ is the volume of Voronoi cell around galaxy $i$.

Next, the code locates the density minima (i.e., the Voronoi cells with the smallest density estimate) and defines these cells to be zone cores. Then, neighboring cells with density estimates that are higher than the zone core are added to the zone. This process is repeated with the new zone boundaries such that any cell with a density estimate larger than a neighboring zone cell is also considered part of that zone. This step ends when no cell with a {larger} density estimate than the cells that comprise the boundary of the zone is found. The result of this step is that there is always a positive density gradient going outwards from a zone core to the boundary of the zone.

Zones {can} then {be} combined to form voids, but there has historically been no consensus on how to control such merging and different methods are known to affect void parameters \citep{nadathur2015c}. {The process of merging involves} adding to a zone all neighboring zones with a minimum Voronoi cell density estimate greater than the minimum density estimate of the starting zone. Like the previous step in the algorithm, this step is repeated for zones on the boundary of the newly constructed void until it can no longer annex additional zones. In order to obtain a usable catalog of non-overlapping voids, it is necessary to introduce some parameters to halt the merging process \citep{nadathur2014b}. 
{Otherwise}, a final ZOBOV void is a collection of all zones of increasing density away from the void center, and excludes the highest-density ridge that separates it from other voids. This produces a hierarchy of voids {and sub-voids}, ranging from many smaller voids to one void that encompasses the entire simulation volume. {Methods to halt merging include using the $P(r)$ statistic (described in more detail below), which measures the probability that any given underdense region is a relic of Poisson noise \citep{neyrinck2008}, as a means to gauge when the growth of a void should be halted. Another strategy is to grow voids until the density of the most underdense Voronoi cell along the ridge separating it from a shallower potential is greater than 0.2 times the mean tracer density of the population \citep{sutter2015}. However, as \cite{nadathur2015c} point out, preventing merging provides a definition of voids that rely solely on the topology of the density field, simplifying the model.} As such, \texttt{REVOLVER} performs no merging and instead defines a void as any zone that has a volume greater than the median value of all zones (see, e.g., \citealt{nadathur2014b}).



As a final pruning step, we compute a cumulative probability that a \texttt{REVOLVER} void arose due to Poisson fluctuations. Since the distribution of underdensities in a Poissonian density contrast field is unknown, \cite{neyrinck2008} calculated the cumulative probability using Monte Carlo sampling of Poissonian data. \cite{neyrinck2008} defined the cumulative probability function, $P(r)$, of the ratio, $r$, between the density of the lowest-density Voronoi cell on the ridge connecting a void to a deeper neighboring void and the minimum density of the void. This ratio is always greater than $1$ (i.e., the ridge of a void region will always be more dense than the center of the void). For the deepest void in a sample, the lowest-density Voronoi cell on the ridge connecting it to its deepest neighbor is used to calculate $r$. In three dimensions, $P(r)$ is given by

\begin{equation}
    \label{eq:p(r)}
    P(r)=\exp[-5.12(r-1)-0.8(r-1)^{(2.8)}] \; .
\end{equation}

\noindent \cite{neyrinck2008} interprets this as the fraction of voids in a Poisson distribution with density contrast ratios greater than $r$. That is, $P(r)$ is the likelihood that an underdense region with ratio $r$ could arise due to Poison noise, with smaller values of $P(r)$ indicating a higher likelihood that the void is spurious. This distribution resembles a half-normal distribution that is centered on $r=1$ with a peak of $P(r)=1$ (see Figure~2 of \citealt{neyrinck2008}). Thus, $P(r)$ has ``standard deviations'' analogous to standard deviations of Gaussian distributions. 
For example, at $1\sigma$ significance, $31.7\%$ of voids have a density contrast ratio greater than $1.22$, and $68.3\%$ of voids have a density contrast ratio less than $1.22$. For our voids, we adopt a significance level of $2\sigma$. That is, we reject all voids with $P(r)<4.55\times10^{-2}$. {This significance level was chosen to be consistent with other studies that have used this technique (see, e.g., \citealt{mao2017}). A lower threshold would admit more voids with low-density cores, while a higher threshold would further restrict our catalogs to only contain voids that show the clearest contrast between core density and linking density. The latter of these thresholds would have reduced our void counts to $\sim20-50\%$ of their current values. A $2\sigma$ cut also ensures that most ($\sim95\%$) voids in the sample have central density contrasts $\lesssim -0.8$ and thus strikes a balance between void counts and removing artifacts of Poisson noise.}

\section{Void Catalogs} 
\label{sec:voidcats}

The results of applying ZOBOV to the galaxy fields of each snapshot are summarized in Table~\ref{tab:voidstats}. Here, {columns 1-4 show the redshift, snapshot number, look back times, and softening lengths for the 11 snapshots of interest, and} columns 5-11 show the number of galaxies that meet the resolution criteria from \S\ref{sec:TNG300snaps}, {the resulting tracer number density ($n_t$)}, the total number of voids found by ZOBOV, the number of voids that meet our void significance threshold, the percent of the volume of the simulation occupied by significant voids{, the number of ``voids-in-clouds'', and the number of ``voids-in-voids'' (see \S\ref{sec:voidradprofs:integrated}).} 
On average, significant voids occupy $33.7\%$ of the simulation volume.


\begin{figure}[!htbp]
    \centering
    \includegraphics[width=\textwidth]{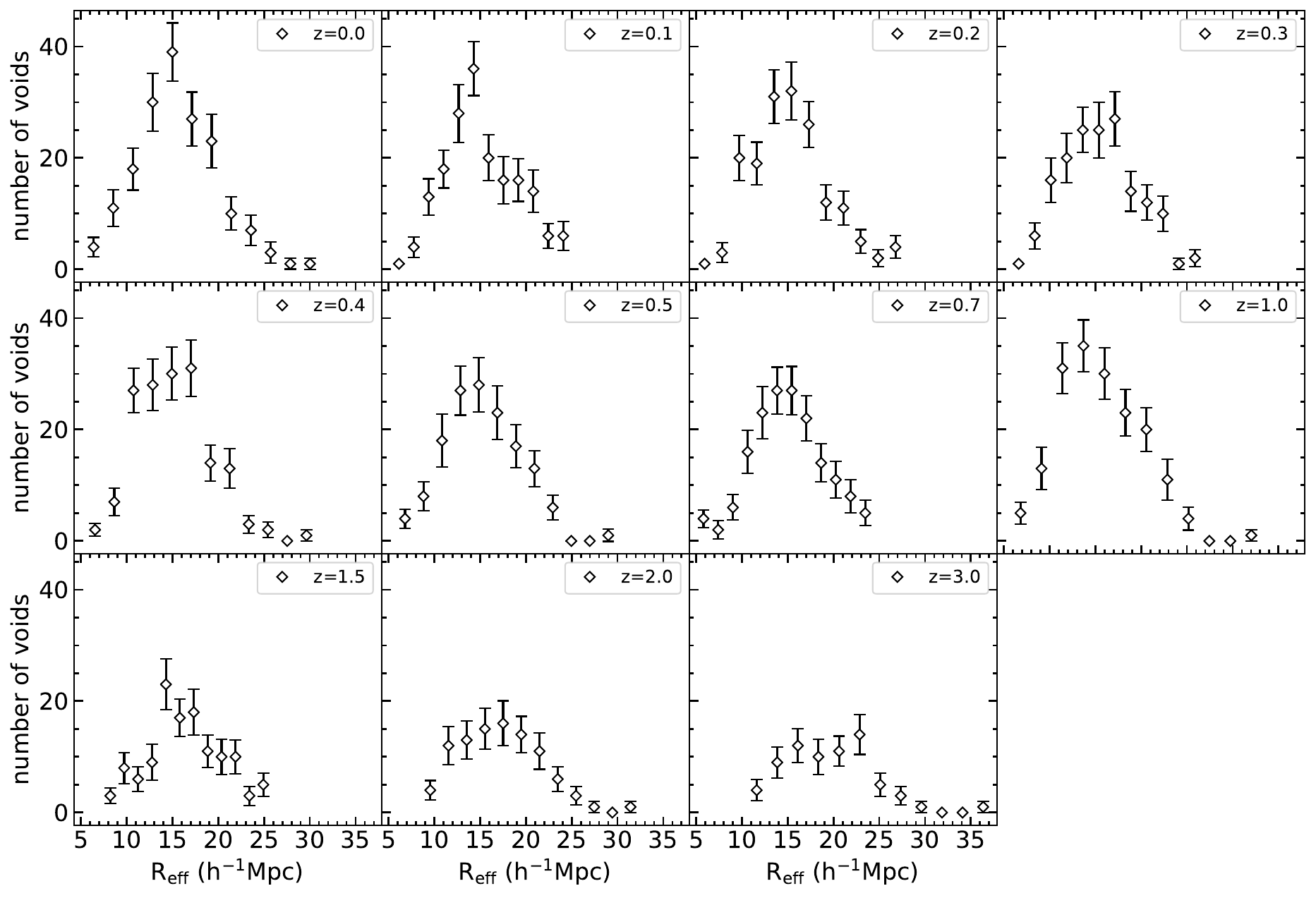}
    \caption{Distributions of void effective radii, $r_{\rm v}$, for each snapshot. {Error bars are calculated from 10,000 bootstrap resamplings of the data.}}
    \label{fig:voidradmosaic}
\end{figure}


The average number of significant voids per snapshot is $147\pm11$. For the most part, the total number of significant voids increases over time. At the earliest redshifts, there is an increase from $70$ significant voids at $z=3.0$ to $123$ at $z=1.5$, then at lower redshifts $\sim145-175$ significant voids are identified in each snapshot. The smaller number of voids identified at the earliest redshifts is likely attributable to the fact that fewer galaxies exist in these snapshots, leading to the identification of fewer voids at earlier redshifts.  Since the galaxy field is sparser at high redshifts than it is at low redshifts, this also results in the high redshift voids being systematically larger on average than the low redshift voids (see below). 


We begin our analysis by computing the effective radii, $R_{\rm eff}$, for the significant voids.  Here, $R_{\rm eff}$ is defined to be the radius of a sphere whose volume is identical to that of the (amorphously shaped) ZOBOV void. Since the voids were identified in co-moving coordinates, the values of $R_{\rm eff}$ are co-moving radii. Figure~\ref{fig:voidradmosaic} shows a mosaic of the distributions of void effective radii from all eleven snapshots. {Each panel shows the distribution of all significant voids in a given snapshot in bins of widths $2h^{-1}$ Mpc.} 
From Figure~\ref{fig:voidradmosaic}, the void effective radii are roughly normally distributed. Over time, the maxima of the distributions increase to just under $20$ between $z=3.0$ and $z=1.5$, where they remain throughout the rest of the snapshots. The maxima of the distributions for redshifts $z\leq1.0$ all occur at $R_{\rm eff} \sim15h^{-1}$Mpc.

\begin{figure}[!htbp]
    \centering
    \includegraphics[width=\textwidth]{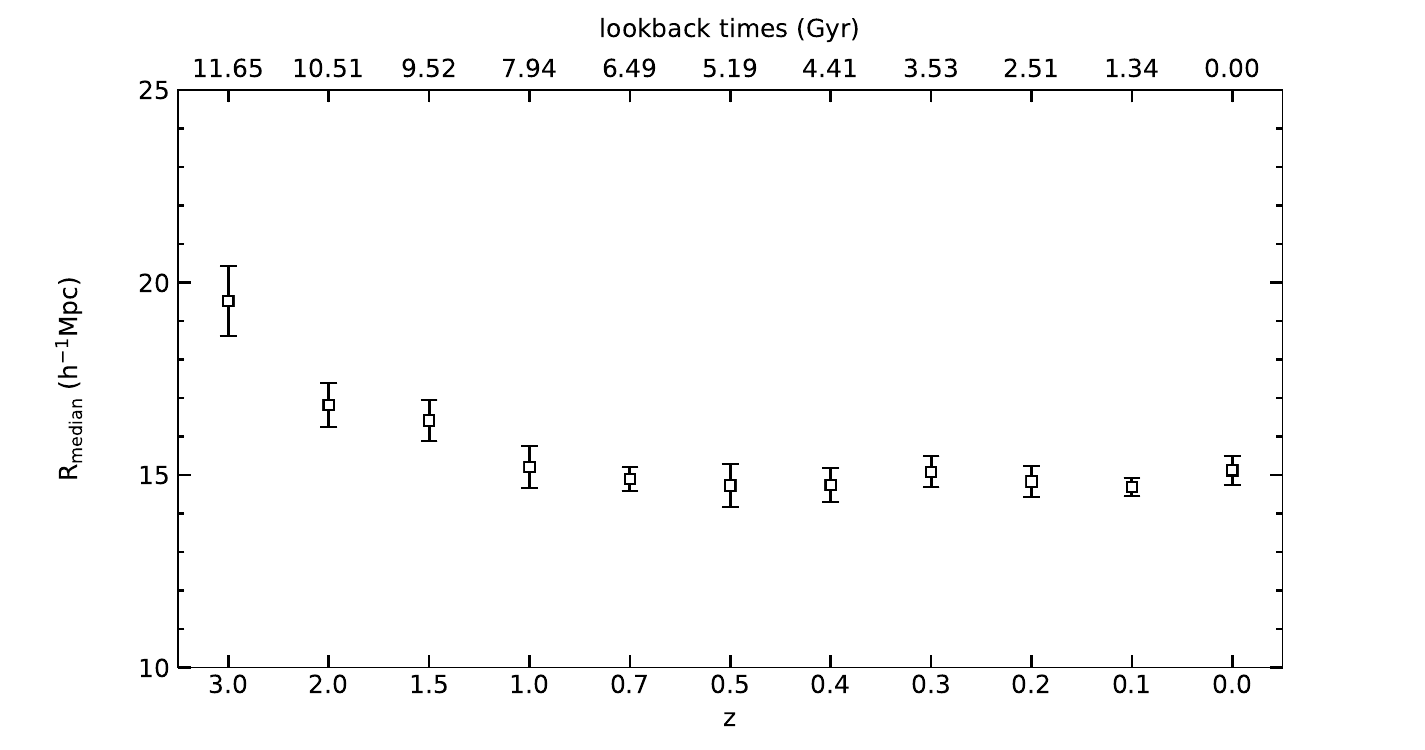}
    \caption{Median $R_{\rm eff}$ of \texttt{TNG300} voids as a function of redshift and lookback time. {Error bars are calculated from 10,000 bootstrap resamplings of the data.} }
    \label{fig:medianvoidradoverz}
\end{figure}

Figure~\ref{fig:medianvoidradoverz} shows the median values of $R_{\rm eff}$ as function of redshift and lookback time. The median effective void radius decreases from $\sim20h^{-1}$Mpc at $z=3.0$ to $\sim15h^{-1}$Mpc at $z=1.0$, where it remains for the rest of the snapshots. As discussed above, the fact that, on average, voids at earlier redshifts are systematically larger than they are at lower redshifts is due to the fact that there are fewer galaxies in the earliest snapshots. In reality, individual voids should increase in size over time as they expand and merge with surrounding voids \citep{sheth}. However, since we are constrained by the resolution of the \texttt{TNG300}, we are limited by the number of galaxies we can use to identify the voids.

\section{Evolution of Void Radial and Integrated Profiles}
\label{sec:voidradprofs}

\subsection{Radial Density Profiles}
\label{sec:voidradprofs:radial}

As must be done in observational studies, we identified \texttt{TNG300} voids using the spatial distributions of the galaxies.  In order to investigate the degree to which light traces mass within the voids, however, we use both the galaxies and the dark matter particles to trace the radial density profiles. Below we will refer to the density profiles as traced by the galaxies as ``luminous matter profiles'' and the density profiles as traced by the dark matter particles as ``dark matter profiles''. {We define a void center as the center of the largest sphere within a void that is completely devoid of galaxies (see, e.g., \citealt{nadathur2015}; \citealt{nadathur2016}; \citealt{nadathur2019}). This was chosen over a volume-weighted barycenter following \cite{nadathur2015} who argued that, compared to a barycentric definition, this minimum density definition gives a better estimate for the true location of the matter underdensity within a void.} We define the radial density profiles as the number density contrast of, respectively, the galaxies and the dark material particles that are contained within radial bins, centered on each void center. That is, the radial dependence of the density contrast (i.e., the void-tracer cross-correlation function) is defined as 

\begin{equation}
    \label{eq:densitycontrast}
    \delta = \frac{n(r)}{\Bar{n}} - 1 .
\end{equation}

\begin{figure}[!htbp]
    \centering
    \includegraphics[width=\textwidth]{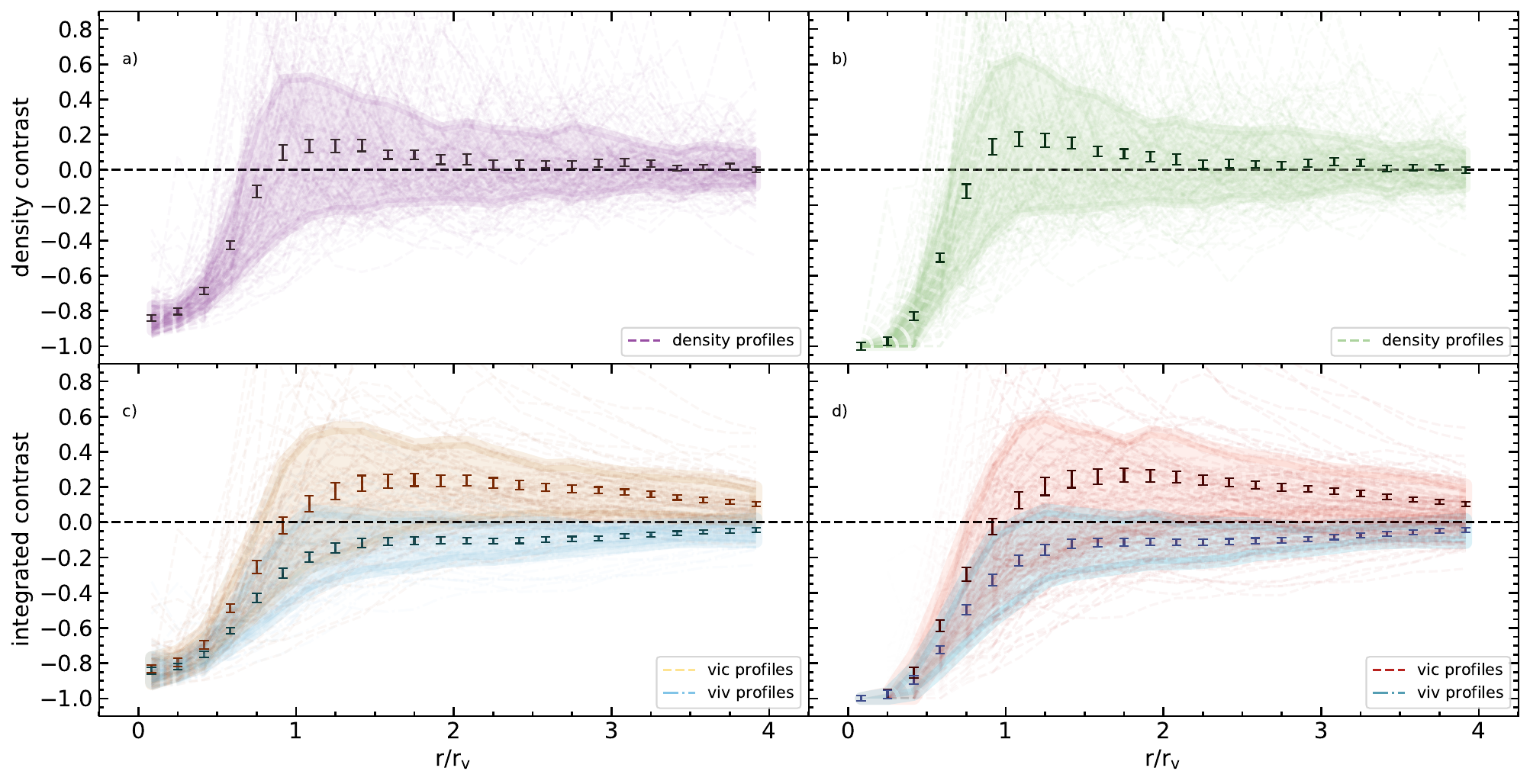}
    \caption{{Density contrast profiles (top) and integrated density profiles (bottom) for all significant voids in the $z=0.0$ snapshot. Panel a): dark matter density contrast profiles (dashed purple lines). Panel c): integrated dark matter density profiles for the ``void-in-cloud'' (dashed orange lines, see text) and ``void-in-void'' (dot-dashed blue lines) populations. Panel b): galaxy density contrast profiles (green). Panel d) integrated galaxy density profiles for the ``void-in-cloud'' (red) and ``void-in-void'' (blue) populations. The error bars represent the standard error of the mean data and the shaded regions show the middle $68\%$ range of the data.}}
    \label{fig:dataspread}
\end{figure}

{Figure~\ref{fig:dataspread} shows the density contrast profiles (top) as well as the integrated number density contrast profiles (bottom) for all voids in the $z=0.0$ snapshot (dashed lines). The left panels use dark matter particles as the tracer population while the right panels use galaxies. Here, the error bars represent the standard error of the mean data while the shaded regions show the middle $68\%$ range of the data. For the integrated density profiles, we classify voids as ``void-in-cloud'' if they are embedded within a larger overdense region and as ``void-in-void'' if they are embedded within a larger underdense region (see \S\ref{sec:voidradprofs:integrated}). 
The range of individual density profiles closely matches the variety of density profiles seen in similar studies (see, e.g., \citealt{paz2013}; \citealt{davila2023}). Most deviations arise within void ridges where the density contrasts can vary between $\sim-0.6-1.5$. Beyond the ridge, the density profiles tend toward the average of the surrounding environment. For most voids, this is a density contrast of $0$, but the profiles for ``voids-in-clouds'' can tend toward values $>0$ at large distances due to the fact that these voids are embedded in an overdense region by definition. 
Despite our small sample size and the range of our individual void profiles, the mean void profiles are relatively localized and thus serve as a good metric for studying the distribution of material in and around voids.} 

\begin{figure}[!htbp]
    \centering
    \includegraphics[width=\textwidth]{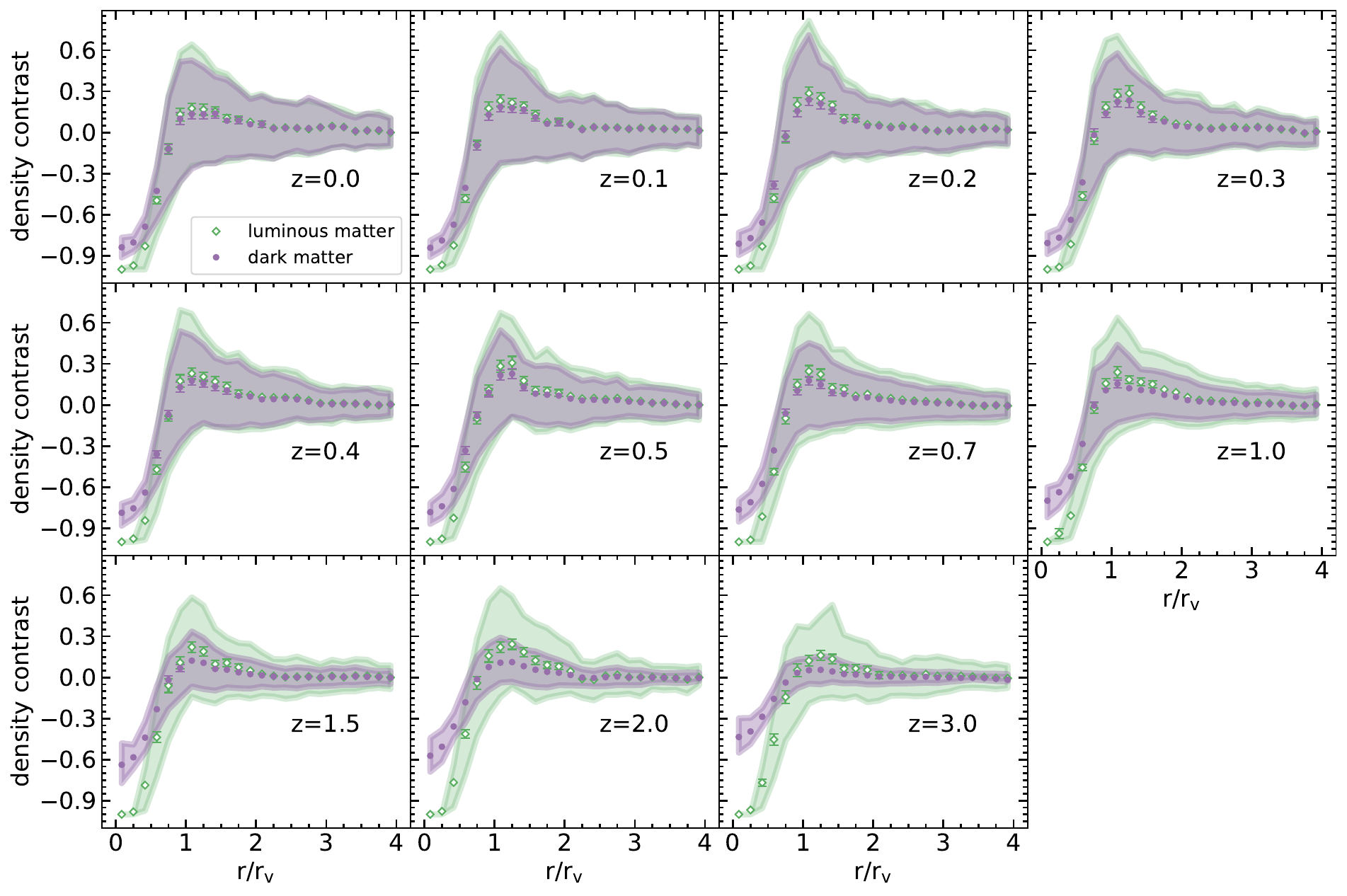}
    \caption{Average radial number density profiles of \texttt{TNG300} voids across time. {Green} diamonds: galaxies are used as density tracers. {Purple} circles: dark matter particles are used as density tracers. {Error bars: standard error of the mean density profile. Shaded regions: middle 68$\%$ range of the data.}} 
    \label{fig:radprofsmosaic}
\end{figure}

Figure~\ref{fig:radprofsmosaic} shows average void radial density profiles for each of the eleven snapshots. {Green} diamonds show the results obtained using galaxy counts and {purple} circles show the results obtained from the dark matter particles. {The error bars show the standard errors of the mean profiles and the shaded regions show the middle $68\%$ ranges of the data.} As expected, these density profiles resemble the reverse spherical top-hat distribution at all redshifts. {While the interiors of voids remain nearly devoid of galaxies ($\delta <-0.9$) across all redshifts, the void centers are more underdense in dark matter particles at later times than they are at earlier times. Despite having $\lesssim123$ voids at redshifts $z\leq1.5$, the dark matter density profiles maintain a relatively narrow range of density contrasts. Over time, the dark matter exits void cores and is deposited into void ridges, increasing the diversity of the individual dark matter profiles such that the ranges of the dark matter profiles become more comparable to those of the galaxy profiles at later times. The ranges of the galaxy profiles also broaden with time, especially at radii $\gtrsim 1r_v$, but the changes are not as noticeable as they are for the dark matter profiles. The mean galaxy profiles do not appear to evolve noticeably with time, although there are slight fluctuations from redshift to redshift that we will discuss below, but the mean dark matter profiles do evolve to have more underdense cores and more overdense ridges at later times. Below, we discuss the evolution of these mean profiles, focusing on how the maximum and minimum density contrasts of each profile evolve over time.}

\begin{figure}[!htbp]
    \centering
    \includegraphics[width=0.9\textwidth, height=0.37\textheight]{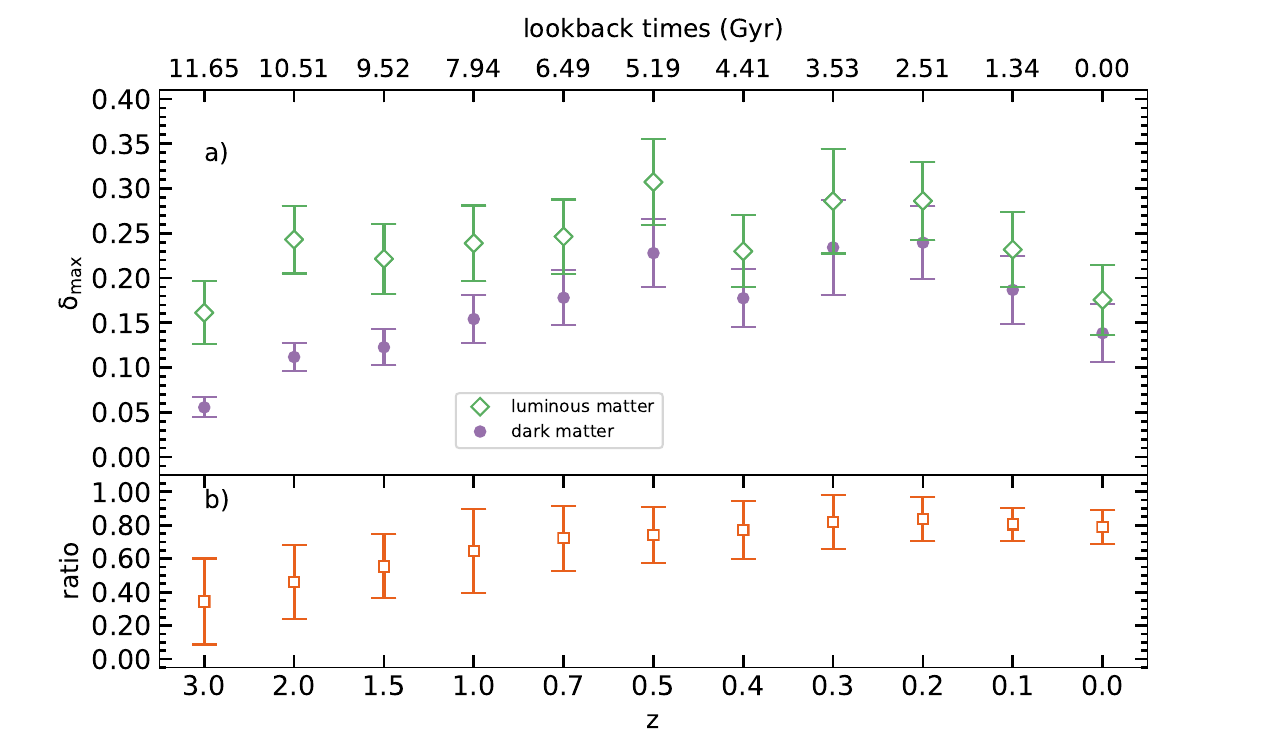}
    \includegraphics[width=0.9\textwidth, height=0.37\textheight]{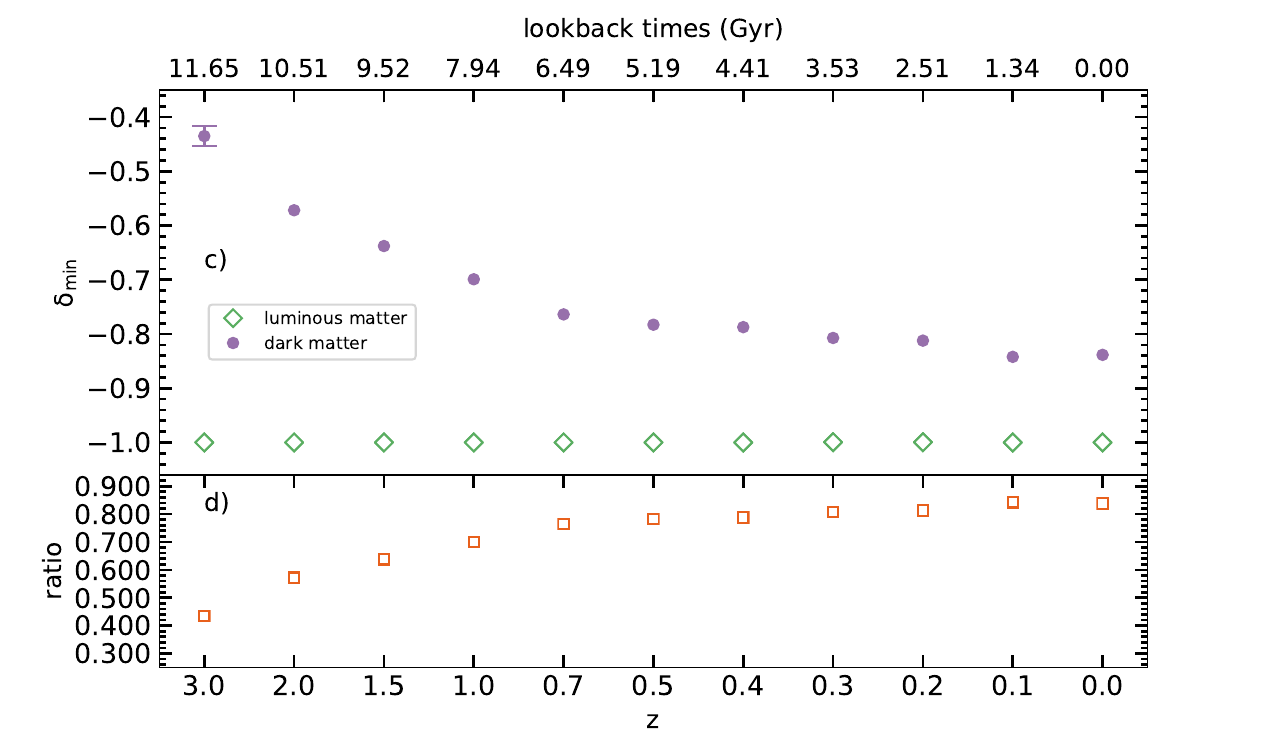}
    \caption{Panels a) and c) show the average maximum and minimum density contrast of all void profiles over time when galaxies (diamonds) and dark matter particles (circles) are used as tracers of the density contrast. Panels b) and d) show the ratios between the corresponding points in panels a) and c).}
    \label{fig:deltaminmax}
\end{figure}

Figure~\ref{fig:deltaminmax} presents summary statistics for the average radial density profiles in Figure~\ref{fig:radprofsmosaic}. Panels a) and c) of Figure~\ref{fig:deltaminmax} show the maximum and minimum density contrasts of the profiles, while panels b) and d) show the ratios {of $\delta_{\rm min,DM}/\delta_{\rm min, galaxies}$} between the corresponding points in panels a) and c). Figure~\ref{fig:deltaminmax}a) demonstrates that the dark matter density contrast of the ridges of voids evolved over time. As traced by the dark matter, the ridges have an average maximum density contrast of $\delta_{\rm max}=0.055\pm0.01$ at $z = 3$. This increases to $0.178\pm0.03$ at $z=0.7$ and $0.240\pm0.041$ $z=0.2$ before decreasing to $0.138\pm0.033$ at $z=0.0$. {The galaxy profiles show a similar trend, increasing from $0.161\pm0.035$ at $z=3.0$ to $0.286\pm0.044$ at $z=0.2$ before decreasing to $0.176\pm0.039$ at $z=0.0$.} 
Because of this, the ratios {of $\delta_{\rm min,DM}/\delta_{\rm min, galaxies}$} in Figure~\ref{fig:deltaminmax}b) slowly increase by a factor of $\sim 2$ from $z=3.0$ to $z=0.0$.

\begin{figure}[!htbp]
    \centering
    \includegraphics[width=0.9\linewidth]{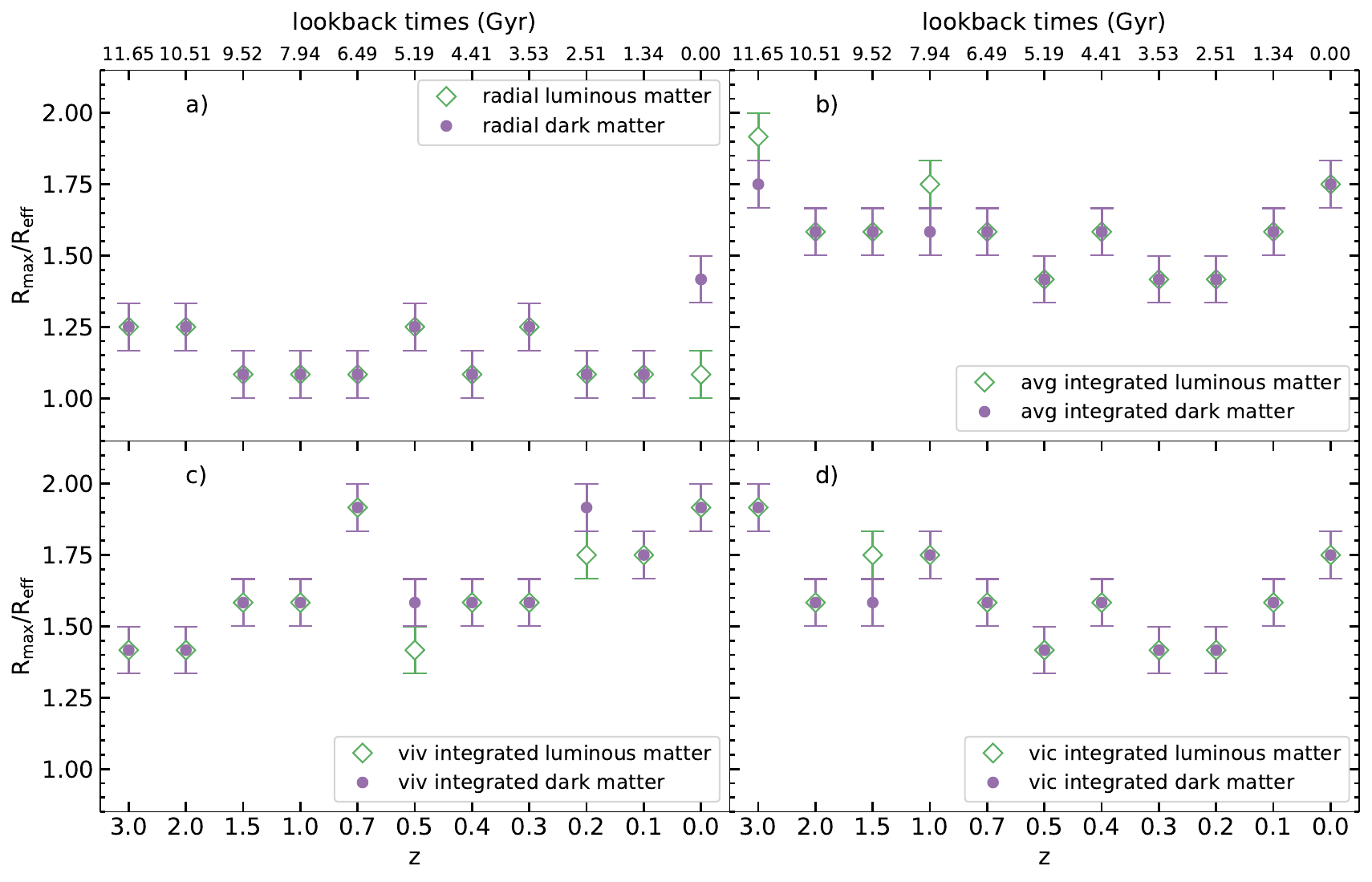}
    \caption{{Panel a): the centers of the radial shells at which $\delta=\delta_{\rm max}$ for the radial density profiles when galaxies (diamonds) and dark matter particles (circles) are used as tracers. Vertical bars show the bin widths used when creating Figure~\ref{fig:radprofsmosaic}. Panels b)-d): the radii of the maximum-density spheres of the integrated density profiles over time (see text). Vertical bars show the bin widths used in constructing Figure~\ref{fig:integratedprofsmosaic}. Panel b): all integrated density profiles. Panel c): ``voids-in-voids.'' Panel d): ``voids-in-clouds.''}}
    \label{fig:Reffmax}
\end{figure}

{An exception to this is an apparent decrease of $\delta_{\rm max}$ between redshifts $z=0.2-0.0$. An evolution of density profiles could indicate a shift of the maximum density to slightly larger radii. To investigate this, we plot $R_{\rm max}$ (i.e., the center of the radial shell at which $\delta=\delta_{\rm max}$) in Figure~\ref{fig:Reffmax}a). Here, green diamonds and purple circles show $R_{\rm max}$ for the galaxy profiles and dark matter profiles, respectively, and the error bars show the radial bin widths of $\sim0.08R_{\rm eff}$. $R_{\rm max}$ remains within 1 radial bin of $1.25\pm0.08R_{\rm eff}$ for all profiles over time. With the exception of the $z=0.0$ snapshot, $R_{\rm max}$ aligns for both the galaxy and dark matter profiles. While we do see an increase in $R_{\rm max}$ for galaxies at $z=0.0$, we see no indication that the densest shells surrounding voids evolve to larger radii at later times. However, as we will see in \S\ref{sec:voidradprofs:integrated}, the integrated density profiles do suggest that $R_{\rm max}$ increases between $z=0.3-0.0$.} 

The average minimum density contrasts in Figure~\ref{fig:deltaminmax}c) occur in the central regions of the voids. Because of this, $\delta_{\rm min}$ represents the density of material in the innermost regions of the voids. Across all redshifts, galaxies are especially rare within the centers of voids; i.e., the luminous matter profiles approach density contrasts of $-1$ at small distances from the void centers. {This could be an artifact of our void sampling as the cut that we imposed with equation~\ref{eq:p(r)} preferentially selects voids with underdense cores relative to their ridges.  
However, despite having nearly empty interiors, Figure~\ref{fig:dataspread}b) shows that even voids with rarefied cores can sharply increase in density as early as $0.3r_v$.} The interiors of voids on the other hand are underdense in dark matter, they are not completely empty of dark matter, and they become increasingly bereft of dark matter over time. From Figure~\ref{fig:deltaminmax}c), most of the changes in the interior dark matter density occur between $z = 3.0$ and $z = 0.7$. That is, at $z=3.0$, the average minimum dark matter density contrast is $-0.42$, while it is $-0.76$ at $z=0.7$, and $-0.82$ at $z=0.0$. Thus, the ratios {of $\delta_{\rm min,DM}/\delta_{\rm min, galaxies}$} in Figure~\ref{fig:deltaminmax}d) increase by a factor of $\sim 2$ from $z=3.0$ to $z = 0.0$.

\begin{figure}[!htbp]
    \centering
    \includegraphics[width=\textwidth]{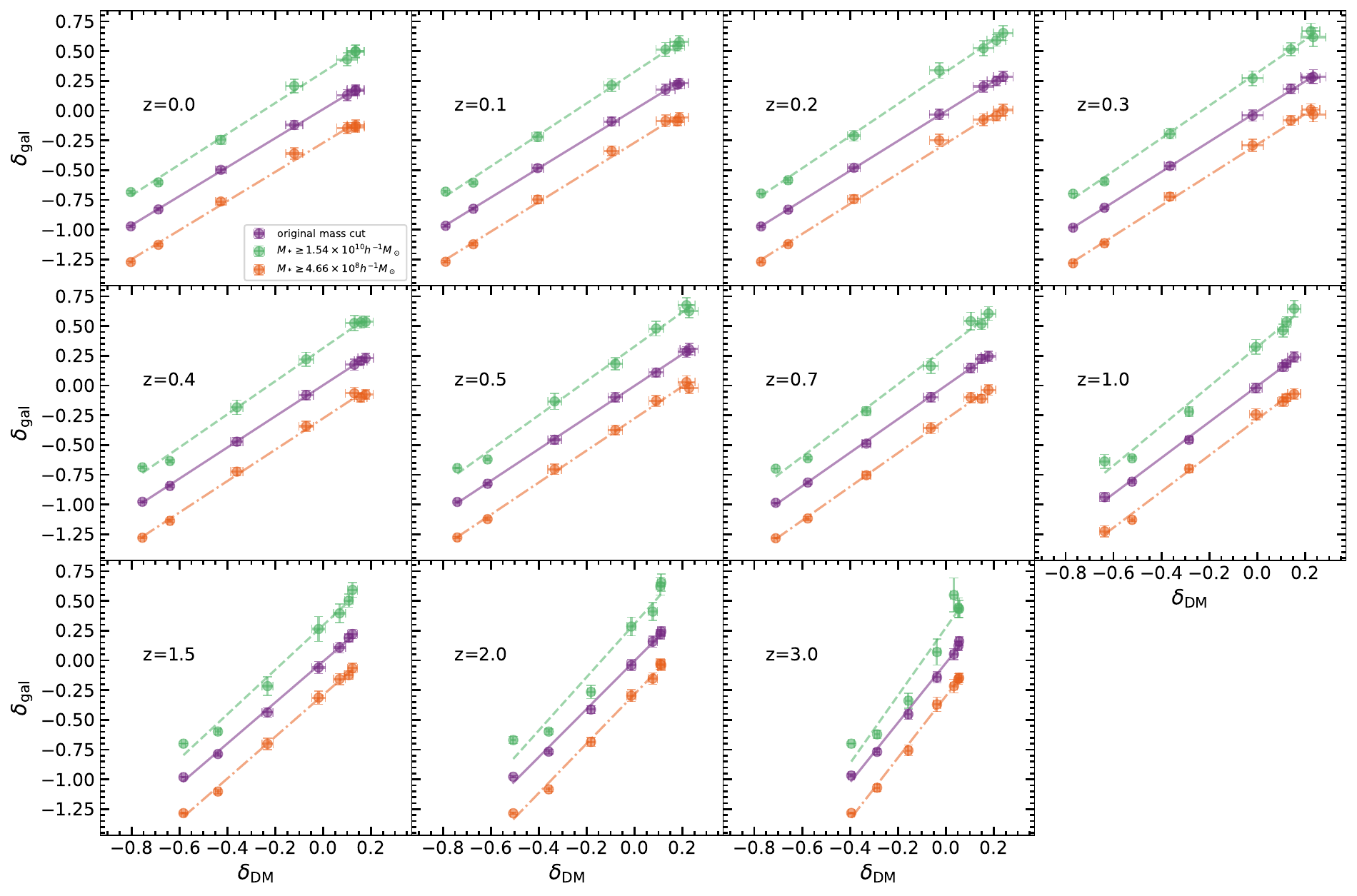}
    \caption{Void galaxy density contrasts vs. dark matter density contrasts in each snapshot. Points: average galaxy and dark matter radial density contrasts in spherical shells with radii between $0.1-1.2R_{\rm eff}$. Minimum density contrasts correspond to void centers; maximum density contrasts correspond to ridges (see Figure~\ref{fig:radprofsmosaic}). Lines correspond to best-fit lines. {Purple: original mass cut (see \S~\ref{sec:voidfinder}). Green: galaxies with stellar masses $\geq 1.54 \times 10^{10}h^{-1}M_\odot$. Orange: galaxies with stellar masses $\geq 4.66 \times 10^{8}h^{-1}M_\odot$. For clarity, the green (orange) points and the green (orange) lines have been linearly shifted up (down) by 0.3.}}
    \label{figbslope}
\end{figure}

\cite{pollina2017} showed that, within voids, luminous tracers of the void density contrast are linearly correlated with the matter density contrast. \cite{pollina2017} found this to be true when galaxies, clusters, or active galactic nuclei were used as tracers. Further, they showed that the slope of this relationship, $b_{\rm slope}$, is a good estimate for the linear bias factor that arises from a consistent shift between the clustering amplitude of the density tracers and the underlying mass field (see, e.g., \citealt{198peebles1980}; \citealt{kaiser1984}; \citealt{mo1996}; \citealt{tinker2010}). That is, they found that the equation

\begin{equation}
    \label{eqb}
    \delta_{\rm{tracer}} = b_{\rm{slope}}\delta_{\rm{DM}} + c_{\rm{offset}} 
\end{equation}

\noindent holds for all voids in the \texttt{Magneticum Pathfinder} simulation at $z=0.14$. Here, $\delta_{\rm tracer}$ is the void tracer density contrast, 
$\delta_{\rm{DM}}$ is the dark matter density contrast, $b_{\rm slope}$ is the slope of the linear relationship, and $c_{\rm offset}$ is the intercept of the relationship (which is usually within a few percent of zero).

\begin{sidewaystable}[hbtp]
    \centering
    \caption{{Best-fitting linear relationships between for equation~\ref{eqb}.}}
    \begin{tabular}{c c c c c | c c c c | c c c c}
         \hline
        & \multicolumn{4}{c|}{original mass cut}  & \multicolumn{4}{c|}{$M_*\geq 4.66\times10^8h^{-1}M_\odot$} & \multicolumn{4}{c}{$M_*\geq 1.54\times10^{10}h^{-1}M_\odot$} \\ 
         \hline
         Redshift & $N_{\rm gal}$ & $b_{\rm slope}$ & $c_{\rm offset}$ & $R^2$ & $N_{\rm gal}$ & $b_{\rm slope}$ & $c_{\rm offset}$ & $R^2$ & $N_{\rm gal}$ & $b_{\rm slope}$ & $c_{\rm offset}$ & $R^2$ \\
         \hline
         0.0 & 342635 & $1.22\pm0.01$ & $0.01\pm0.01$ & 0.999   & 226171 & $1.22\pm0.04$ & $0.03\pm0.02$ & 0.996 & 41450 & $1.29\pm0.03$ & $0.02\pm0.01$ & 0.996  \\
         0.1 & 338602 & $1.23\pm0.01$ & $0.01\pm0.01$ & 0.999   & 222690 & $1.25\pm0.04$ & $0.03\pm0.02$ & 0.996 & 39878 & $1.32\pm0.03$ & $0.03\pm0.01$ & 0.997 \\
         0.2 & 334219 & $1.25\pm0.01$ & $0.00\pm0.01$ & 0.999   & 218070 & $1.27\pm0.03$ & $0.02\pm0.01$ & 0.996 & 38327 & $1.35\pm0.03$ & $0.03\pm0.01$ & 0.998 \\
         0.3 & 329227 & $1.27\pm0.01$ & $-0.01\pm0.01$ & 0.999  & 212056 & $1.28\pm0.03$ & $0.02\pm0.01$ & 0.998 & 36536 & $1.38\pm0.03$ & $0.02\pm0.01$ & 0.997 \\
         0.4 & 325406 & $1.31\pm0.01$ & $0.00\pm0.01$ & 0.999   & 205199 & $1.32\pm0.03$ & $0.02\pm0.01$ & 0.997 & 34609 & $1.39\pm0.04$ & $0.01\pm0.02$ & 0.996 \\
         0.5 & 322657 & $1.33\pm0.01$ & $0.00\pm0.01$ & 0.999   & 197583 & $1.35\pm0.03$ & $0.02\pm0.01$ & 0.997 & 32458 & $1.45\pm0.05$ & $0.03\pm0.02$ & 0.995 \\
         0.7 & 315329 & $1.41\pm0.02$ & $0.00\pm0.01$ & 0.999   & 183228 & $1.42\pm0.03$ & $0.02\pm0.01$ & 0.998 & 28148 & $1.52\pm0.06$ & $0.02\pm0.02$ & 0.991 \\
         1.0 & 295677 & $1.51\pm0.02$ & $-0.01\pm0.01$ & 0.999   & 163949 & $1.52\pm0.05$ & $0.02\pm0.02$ & 0.995 & 22058 & $1.66\pm0.09$ & $2.46\pm0.03$ & 0.986 \\
         1.5 & 229435 & $1.72\pm0.04$ & $-0.01\pm0.01$ & 0.998  & 121871 & $1.76\pm0.04$ & $0.01\pm0.01$ & 0.998 & 14210 & $1.87\pm0.10$ & $-0.01\pm0.03$ & 0.985 \\
         2.0 & 170014 & $2.02\pm0.06$ & $0.00\pm0.02$ & 0.996   & 84488  & $2.08\pm0.07$ & $0.02\pm0.02$ & 0.995 & 8920  & $2.24\pm0.21$ & $0.01\pm0.05$ & 0.957 \\
         3.0 & 118841 & $2.51\pm0.09$ & $-0.02\pm0.02$ & 0.994   & 35505  & $2.59\pm0.08$ & $0.00\pm0.02$ & 0.996 & 2952  & $2.85\pm0.32$ & $-0.02\pm0.06$ & 0.941\\
         \hline
    \label{tabbparams}
    \end{tabular}
\end{sidewaystable}

To determine whether our \texttt{TNG300} voids show a similar linear relationship, {we plot $\delta_{\rm gal}$ vs $\delta_{\rm DM}$ using the data reported above in Figure~\ref{figbslope} (plotted in purple).} In addition, to facilitate a comparison with the results of \cite{illustris4} below, {we recalculate our void-galaxy cross-correlation functions using as tracer populations only the galaxies with stellar masses $\geq 4.66\times10^8h^{-1}M_\odot$ and $1.54\times10^{10}h^{-1}M_\odot$, and we plot these data in Figure~\ref{figbslope} in orange and green, respectively. For clarity, the green (orange) points and line have been linearly shifted up (down) by 0.3.} Figure~\ref{figbslope} shows $\delta_{\rm gal}$ vs. $\delta_{\rm DM}$ for the average void profile from each snapshot, and the dashed lines are best-fit linear relationships. That is, each point in Figure~\ref{figbslope} shows the galaxy and dark matter radial density contrasts in concentric spherical shells with radii between $0.1-1.2R_{\rm eff}$, such that the minimum density contrasts correspond to void centers and the maximum density contrasts correspond to ridges (see Figure~\ref{fig:radprofsmosaic}). A linear trend between $\delta_{\rm gal}$ and $\delta_{\rm DM}$ exists at all redshifts {for all 3 tracer populations}, and the slope of the relationship generally increases between sequential snapshots. Below, we show that this increasing slope indicates a larger clustering bias between galaxies and dark matter at high redshifts. Table~\ref{tabbparams} shows the number of tracers, $b_{\rm slope}$, and $c_{\rm offset}$ for each snapshot, along with {the $R^2$ statistics for the fits. Here, the left columns correspond to the results when galaxies in the original mass cut are used as tracers (i.e., galaxies with stellar masses $\geq10^8h^{-1}M_\odot$ at $z<3.0$ and $\geq10^{7.75}h^{-1}M_\odot$ at $z=3.0$), while the center (right) columns show the results when galaxies $\geq 4.66\times 10^8h^{-1}M_\odot$ ($\geq 1.54\times 10^{10}h^{-1}M_\odot$) are used as traces.}

\cite{illustris4} calculated the clustering bias within \texttt{TNG300} using halos and galaxies as density tracers.  These were selected based on several resolution cuts to the virial mass, stellar mass, instantaneous star formation rates, and specific star formation rates. 
%
{The resolution cuts imposed by \cite{illustris4} correspond to galaxy (or halo) densities that cover the ranges of ongoing and upcoming surveys. They perform a series of mass cuts such that the number density of tracers within each snapshot corresponds to $0.03h^3\rm{Mpc}^{-3}$ or $0.003h^3\rm{Mpc}^{-3}$. At $z=0.0$, this corresponds to a stellar mass cut of $4.66\times 10^{8}h^{-1}M_\odot$ and $3\times10^{-2}h^3\rm{Mpc^3}$, respectively}. {\cite{illustris4} then define a scale-dependent bias model of the form}

\begin{equation}
    b(k) = b_0 + \beta \left(\ln\frac{k}{k_0} \right)^2 ,
\end{equation}

\noindent {where $b_0$ is the large-scale linear bias, $\beta$ is the strength of the scale dependence, $k$ is a wave number, and $k_0=0.02h\rm{Mpc}^{-1}$ is the wave number where $db(k)/dk=0$. Since we fix our tracer number densities to be equal to theirs at $z=0.0$, we can directly compare our $b_{\rm slope}$ values with their reported values of $b(k)$. For galaxies with stellar masses $\geq 4.66\times 10^{8}h^{-1}M_\odot$ ($\geq 1.54\times10^{10}h^{-1}M_\odot$), they fit $b_0=1.17$ ($b_0=1.36$) and $\beta=-0.0007$ ($\beta=0.0016$). If the wave number is $0.067h\rm{Mpc}^{-1}$, which corresponds to the typical size scales of our voids, the value of $b(k)$ is $1.17$ ($1.38$), while we report a corresponding value of $b_{\rm slope}=1.22\pm0.4$ ($b_{\rm slope}=1.29\pm0.03$).} 


When we use galaxies with stellar masses $\geq 4.66\times10^8h^{-1}M_\odot$ as a density tracer, we obtain slopes that are very close to the bias estimates reported by \cite{illustris4}. 
That is, using the same selection criteria as ours {and at typical length scales for our voids}, \cite{illustris4} report a {scale-dependent} bias parameter of $1.17$ at $z=0.0$, {while we report a value of $b_{\rm slope}=1.22\pm0.04$. When we use as a tracer population galaxies with stellar masses $\geq 1.54\times10^{10}h^{-1}M_\odot$, our $b_{\rm slope}$ estimate deviates from the results of \cite{illustris4} ($b(k)=1.38$ vs. $b_{\rm slope}=1.29\pm0.03$).} 


{Had we employed either of these stellar masses during our void identification, then the sparser tracer populations would have increased the average sizes of our reported voids. As \cite{illustris4} note, even at higher redshifts where the bias is more prominent, the bias of the galaxies in \texttt{TNG300} becomes independent on distance scales larger than $\sim1-10h^{-1}$Mpc. As our voids are already larger than this ($\sim15-20h^{-1}$Mpc), we do not expect the bias to change at higher size scales. However, \cite{pollina2017} report that $b_{\rm slope}$, as measured from galaxy voids in the \texttt{Magneticum Pathfinder} simulation that were identified with the \texttt{ZOBOV} algorithm, decreases with increasing void size up to void sizes $\gtrsim 60h^{-1}$Mpc where it becomes independent of void size. Unfortunately, the limited size of the \texttt{TNG300} simulation makes such an analysis impossible even if we had used a sparser tracer population during void identification, so we do not attempt to do so here.}

The relationship between $b_{\rm slope}$ and the bias parameter is expected to occur in the linear regime of density fluctuations. Thus, our results are in agreement with \cite{pollina2017} and \cite{pollina2019}, demonstrating that the assumption of linear bias between the galaxies and dark matter within voids holds for voids defined from the density field. Here, we have shown that this assumption continues to hold out to $z=3.0$.   


\subsection{Integrated Profiles}
\label{sec:voidradprofs:integrated}

In this section, we define two void populations: ``voids-in-voids'' and ``voids-in-clouds'' {using the same selection of galaxies that we used in \S\ref{sec:voidfinder}.} These structures correspond to underdensities in galaxy distributions that are embedded, respectively, within larger underdensities or overdensities (see, e.g., \citealt{sheth}; \citealt{paz2013}).

Using galaxies and dark matter particles as density tracers, we calculate integrated number density contrast profiles, $\Delta(r)$, separately for ``voids-in-voids,'' ``voids-in-clouds,'' and the entire void population.  That is, $\Delta(r)$ includes the total number of tracer objects in concentric spheres of radius, $r$, centered on the void centers. {``Voids-in-clouds'' form from the collapse of larger overdensities, while ``voids-in-voids'' arise from the expansion and merging of underdense regions, meaning the surrounding velocity field is a useful tool to distinguish void type. In the linear regime, the velocity profiles of voids are proportional to $-\Delta(r)$ (\citealt{198peebles1980}; \citealt{peebles1993}), making the integrated density profile a well-established proxy for both} void hierarchy (see, e.g., \citealt{sheth}; \citealt{ceccarelli2013}; \citealt{paz2013}; \citealt{davila2023}; \citealt{curtis2024}) {and void velocity profiles (see, e.g., \citealt{sheth}; \citealt{paz2013}; \citealt{hamaus2014}; \citealt{schuster2023})}. Using the criteria from \cite{sheth} and \cite{ceccarelli2013}, we define ``voids-in-voids'' to be those with integrated galaxy density contrasts $< 0$ at $r=3R_{\rm eff}$ and ``voids-in-clouds'' to be those with integrated galaxy density contrasts $> 0$ at $r=3R_{\rm eff}$. Columns 10 and 11 of Table~\ref{tab:voidstats} list the total number of ``voids-in-clouds'' and ``voids-in-voids,'' respectively. From Table~\ref{tab:voidstats}, there are more ``voids-in-clouds'' than ``voids-in-voids,'' with an average ratio of $4:3$ for each snapshot. 


\begin{figure}[!htbp]
    \centering
    \includegraphics[width=\textwidth]{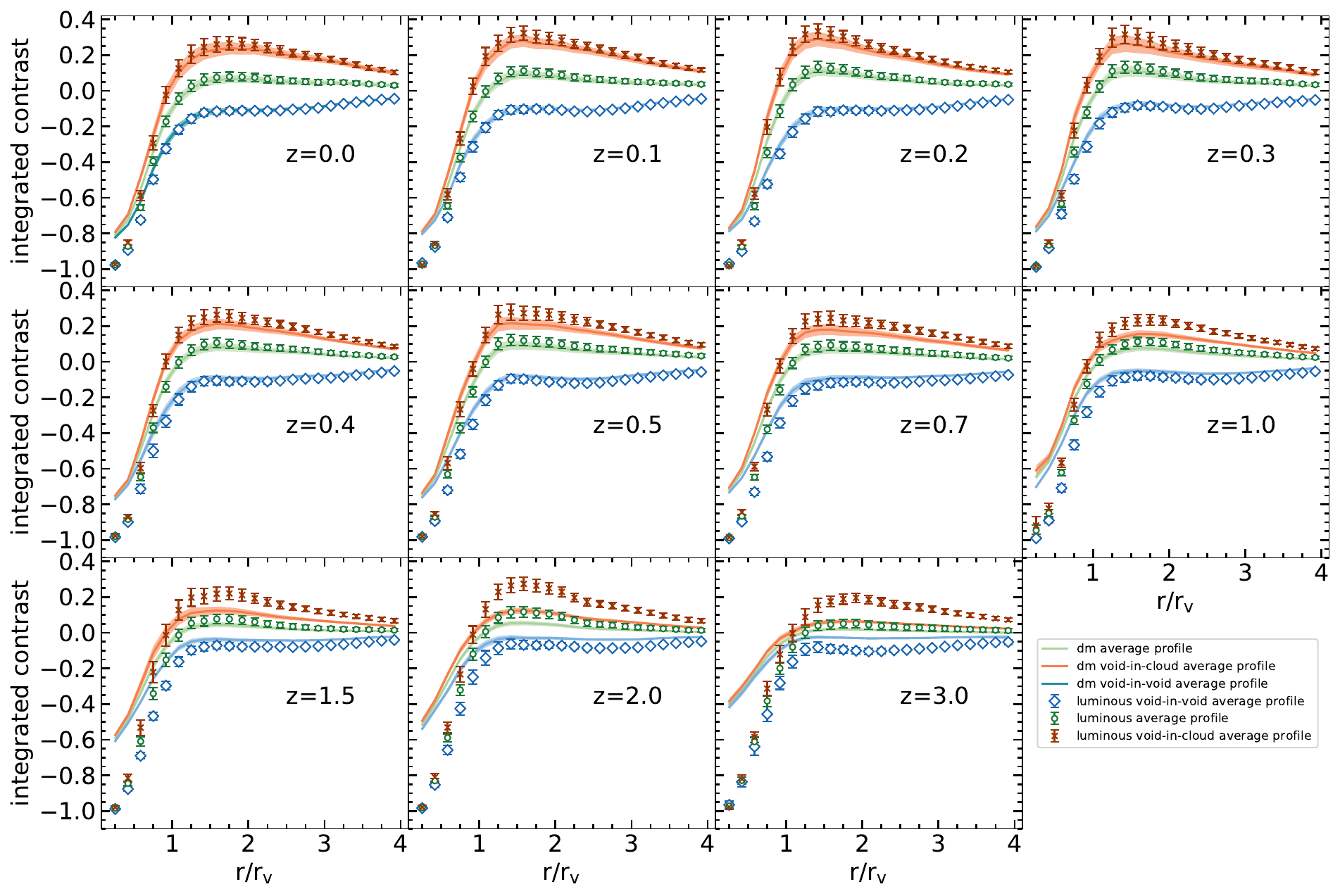}
    \caption{Integrated density profiles for ``voids-in-voids,'' ``voids-in-clouds,'' and the entire void population when galaxies (points) and dark matter particles (lines) are used as density tracers. {Error bars and shaded regions show the standard error of the mean dark matter and galaxy profiles, respectively.} Red points and orange lines: ``voids-in-clouds.'' Green points and green lines: the entire void population. Blue points and blue lines: ``voids-in-voids.''}
    \label{fig:integratedprofsmosaic}
\end{figure}

Figure~\ref{fig:integratedprofsmosaic} shows the average integrated profiles for each void population. Points show the results when galaxies are used as density tracers, and solid lines show the results when dark matter particles are used as density tracers. Red points and orange lines show profiles for ``voids-in-clouds,'' green points and green lines show profiles for the entire void population, and blue points and blue lines show profiles for the ``voids-in-clouds.''

Each distribution resembles a reverse spherical top-hat distribution, and similarities and differences exist between the void populations. First, in all cases, the interiors of each void population are nearly empty of galaxies and they are underdense, but not completely devoid of, dark matter. Secondly, at a given redshift (with the exception of $z=1$), all three luminous profiles converge to nearly identical interior integrated density contrasts, and the same holds true for the dark matter profiles.
However, the integrated density contrasts of the three void populations (for a given tracer) begin to deviate noticeably from each other for $r \gtrsim 0.5R_{\rm eff}$. That is, the ``void-in-cloud'' profiles rapidly increase in luminous and dark matter integrated density, and they peak at integrated density contrasts $\sim0.30$. In contrast, the ``void-in-void'' profiles increase only gradually beyond $r\sim0.5R_{\rm eff}$ and reach density contrasts of $\sim-0.2$. Thirdly, there is a higher integrated density of galaxies than dark matter at the void ridges in the ``void-in-cloud'' and total void populations, whereas dark matter has a larger integrated density contrast than galaxies in the ridges of ``voids-in-voids.'' 

{Lastly, regarding redshift evolution, the integrated dark matter density profiles tend to align more closely with their corresponding integrated galaxy density profiles over time. That is, while the centers of voids remain nearly devoid of galaxies at all redshifts, the dark matter profiles steadily decrease from $\sim-0.4$ at $z=3.0$ to $\sim-0.7$ at $z=0.7$ and $\sim-0.8$ at $z=0.0$. Similarly, the differences between the integrated dark matter and galaxy profiles along void ridges tend to decrease over time. At $z=3.0$, the ridges of ``voids-in-clouds'' have a mean integrated galaxy density of $\sim0.2$ and a mean integrated dark matter density of $\sim0.06$, while at $z=0.2$, the mean integrated galaxy density of these voids is $\sim0.34$ and the integrated dark matter density is $\sim0.29$.}

\begin{figure}[!htbp]
    \centering
    \includegraphics[width=0.9\textwidth, height=0.37\textheight]{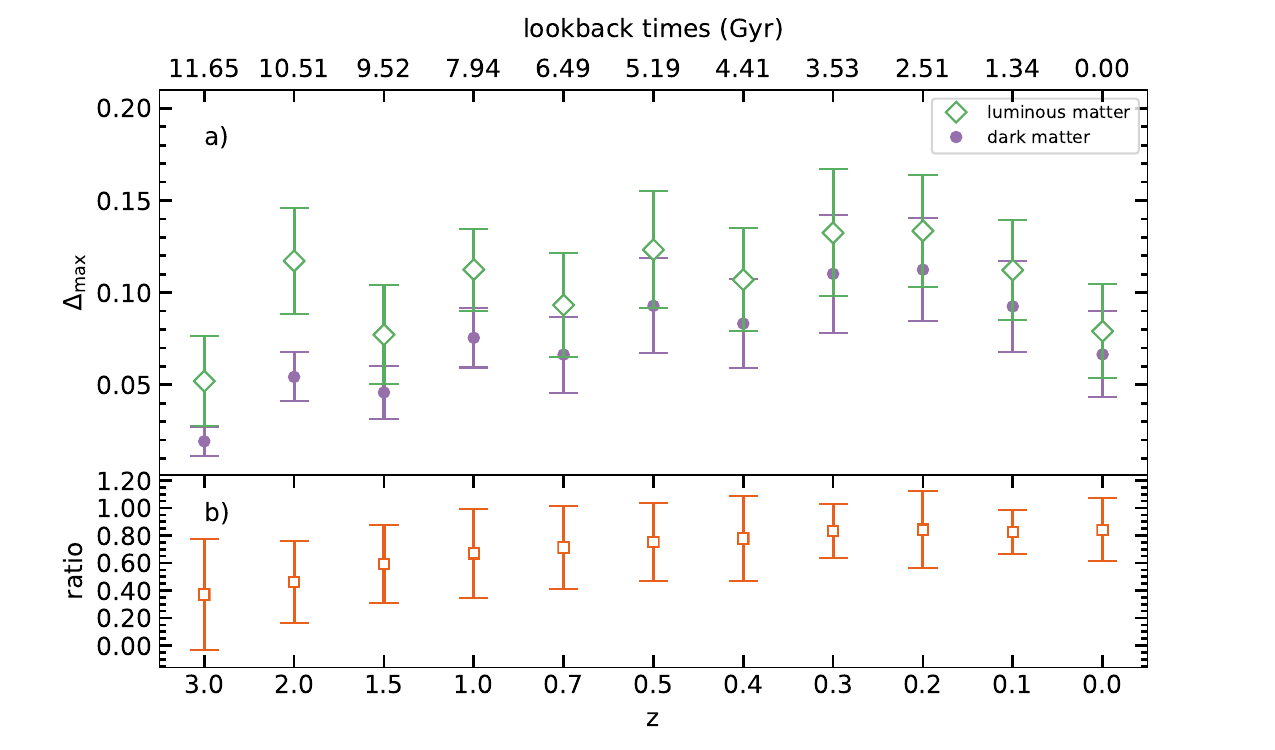}
    \includegraphics[width=0.9\textwidth, height=0.37\textheight]{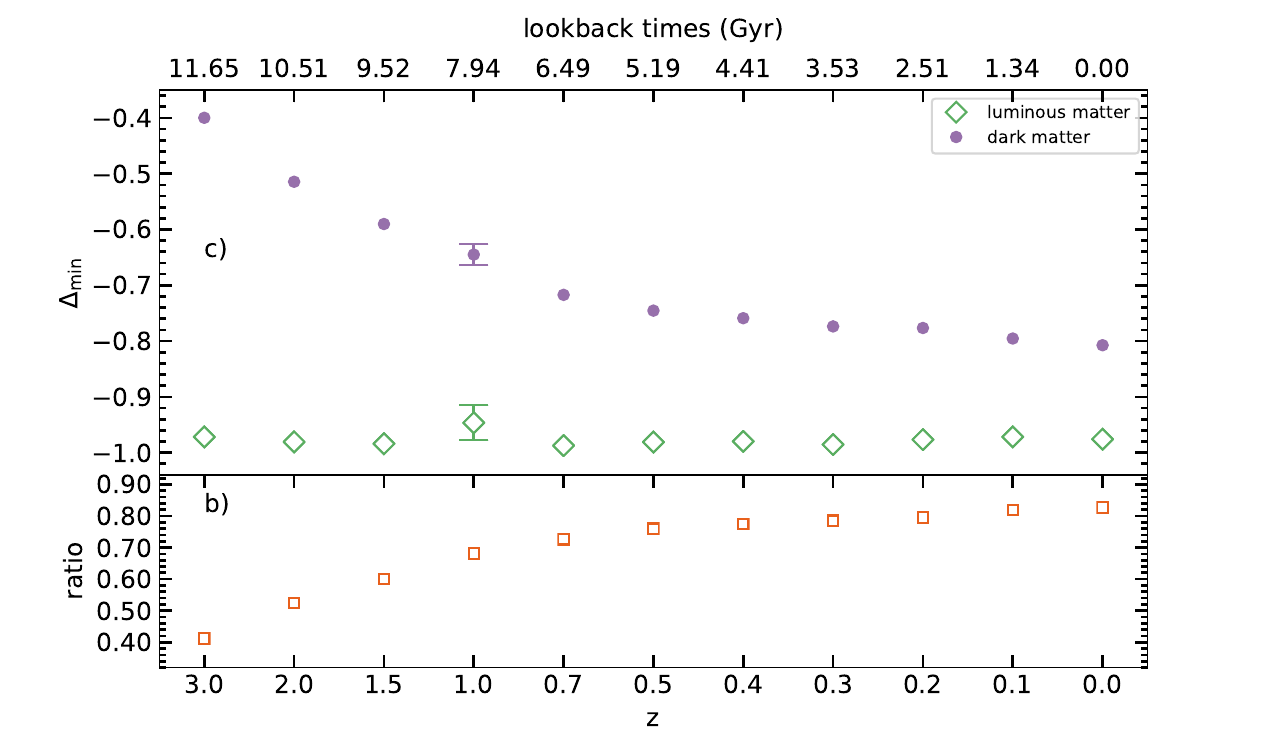}
    \caption{Same as Figure~\ref{fig:deltaminmax} except for all integrated void profiles.}
    \label{fig:integrateddeltaminmax}
\end{figure}

\begin{figure}[!htbp]
    \centering
    \includegraphics[width=0.9\textwidth, height=0.37\textheight]{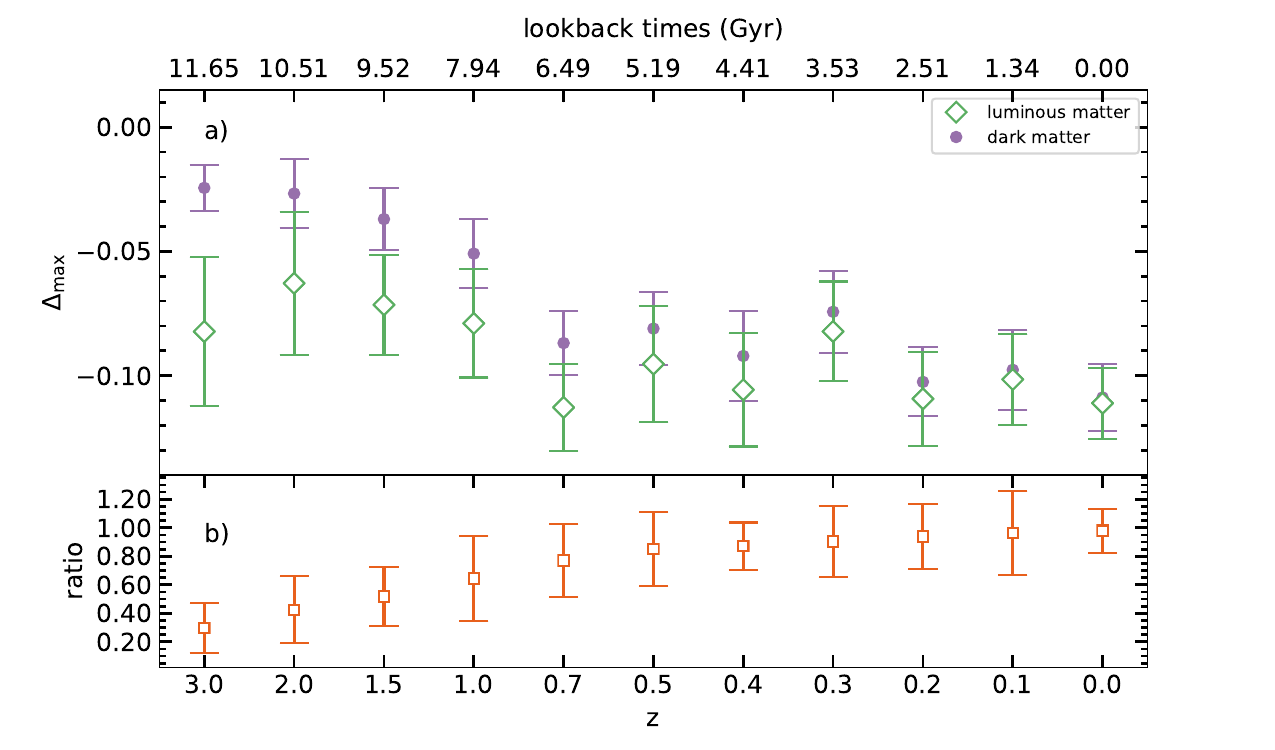}
    \includegraphics[width=0.9\textwidth, height=0.37\textheight]{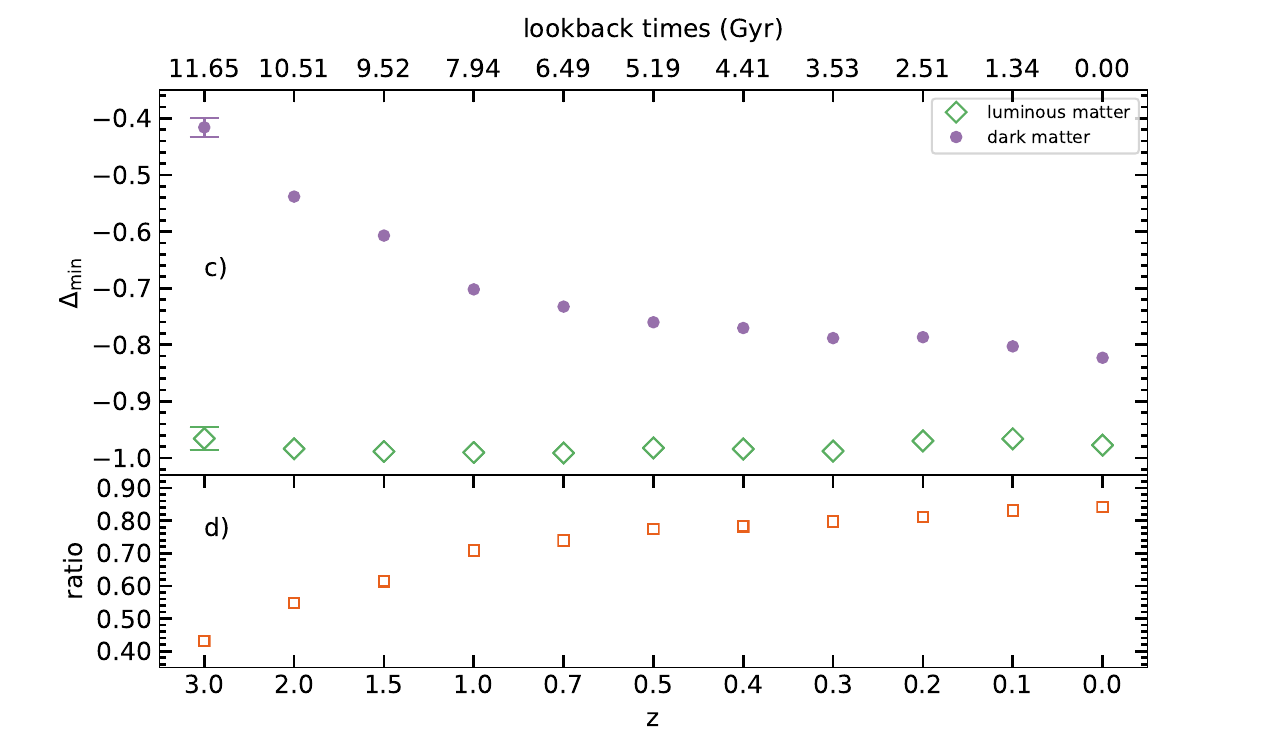}
    \caption{Same as Figure~\ref{fig:deltaminmax} except for all integrated ``void-in-void'' profiles.}
    \label{fig:vivintegrateddeltaminmax}
\end{figure}

\begin{figure}[!htbp]
    \centering
    \includegraphics[width=0.9\textwidth, height=0.37\textheight]{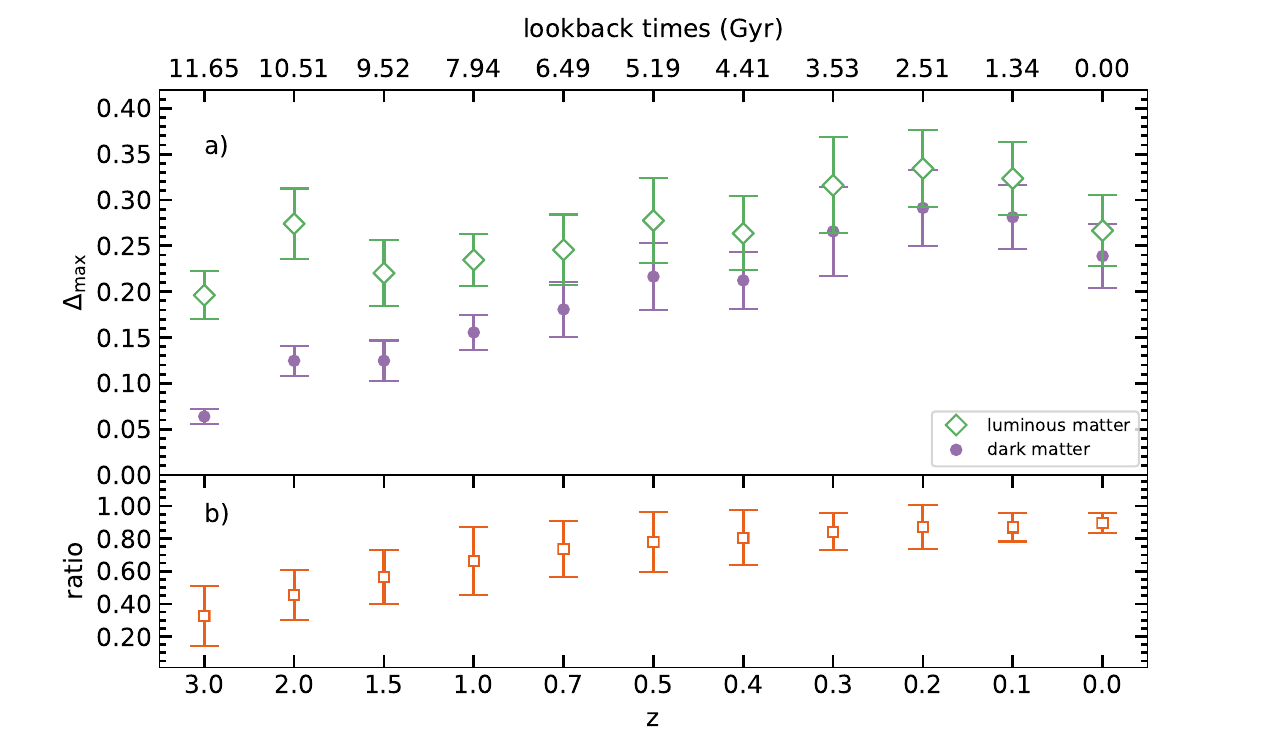}
    \includegraphics[width=0.9\textwidth, height=0.37\textheight]{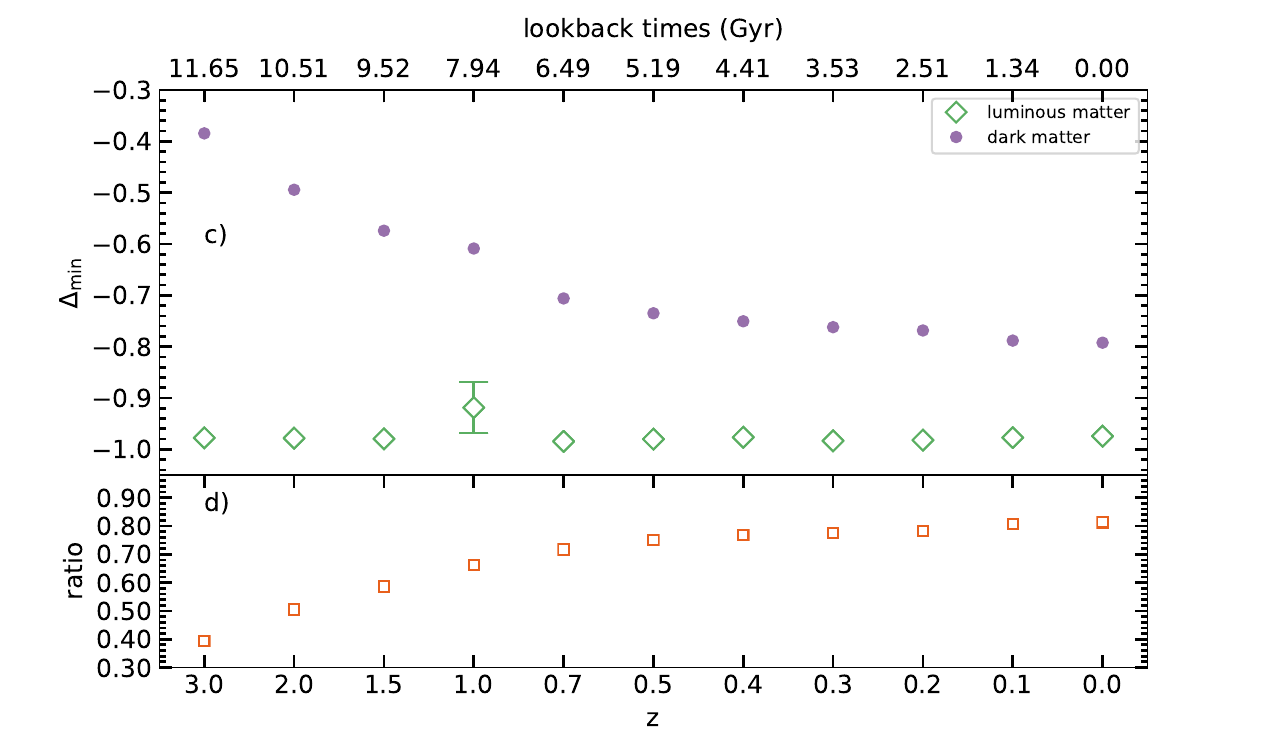}
    \caption{Same as Figure~\ref{fig:deltaminmax} except for all integrated ``void-in-cloud'' profiles.}
    \label{fig:vicintegrateddeltaminmax}
\end{figure}

{We examine the time evolution of these profiles in more detail} in Figure~\ref{fig:integrateddeltaminmax}, Figure~\ref{fig:vivintegrateddeltaminmax}, and Figure~\ref{fig:vicintegrateddeltaminmax}. Each Figure is formatted in the same way as Figure~\ref{fig:deltaminmax}. In discussing these figures, we will refer to the sphere at which the maximum number density contrast occurs as the ``maximum-density sphere'' and the sphere at which the minimum number density contrast occurs as the ``minimum-density sphere.''

Figure~\ref{fig:integrateddeltaminmax}a) shows the integrated density contrasts of maximum-density spheres for the entire void sample. These maximum-density spheres extend to just beyond $R_{\rm eff}$. 
From Figure~\ref{fig:integrateddeltaminmax}a), the maximum-density spheres contain somewhat less dark matter at early redshifts than at $z=0.0$. {That is, $\Delta_{\rm max}$ for the dark matter is $0.019\pm0.007$ at $z=3.0$, $0.046\pm0.014$ at $z=1.5$, $0.112\pm0.028$ at $z=0.2$, and $0.066\pm0.023$ at $z=0.0$.} {The maximum-density spheres for the galaxies fluctuate around $0.052\pm0.024$ at $z=3.0$ to $0.0132\pm0.030$ at $z=0.3$ and $0.079\pm0.025$ at $z=0.0$.} The ratios {of $\Delta_{\rm min,DM}/\Delta_{\rm min, galaxies}$} shown in Figure~\ref{fig:integrateddeltaminmax}b) thus increase over time. Here, the ratios asymptotically approach a value of $0.84\pm0.23$ at low redshifts.

{Figure~\ref{fig:Reffmax}b) shows the radii of the maximum-density spheres ($R_{\rm max}$) used to calculate $\Delta_{\rm max}$ for the integrated profiles described above. Here, diamonds (circles) show $R_{\rm max}$ when galaxies (dark matter particles) are used as tracers and bars show the bin widths used in Figure~\ref{fig:integratedprofsmosaic}. With the exception of the $z=3.0$ and $z=1.0$ snapshots, the values of $R_{\rm max}$ for the dark matter profiles are equal to those calculated for the galaxy profiles. At redshifts $z\leq2.0$, $R_{\rm max}$ remains within one radial bin of $1.58\pm0.08R_{\rm eff}$. For both the dark matter and the galaxy profiles at $z=0.2$, $R_{\rm max}$ steadily increases from $1.42\pm0.08R_{\rm eff}$ to $1.58\pm0.08R_{\rm eff}$ at $z=0.1$ and $1.75\pm0.08R_{\rm eff}$ at $z=0.0$. This correlates with the decrease in $\Delta_{\rm max}$ that we report for these redshifts in Figure~\ref{fig:integrateddeltaminmax}, indicating a mean shift of $R_{\rm max}$ to slightly larger radii at lower redshifts.}

Figure~\ref{fig:integrateddeltaminmax}c) shows the density contrasts in minimum-density spheres. These minimum-density spheres all have radii $r\leq0.25R_{\rm eff}$. They are almost completely devoid of galaxies across all redshifts and consistently have galaxy number density contrasts of $\sim-0.98$. This is not the case for the minimum-density spheres for the dark matter, which have density contrasts that decrease significantly over time. For instance, {the integrated dark matter profile decreases from $\Delta_{\rm min}=-0.400\pm0.012$ at $z=3.0$, to $\Delta_{\rm min}=-0.807\pm0.007$ at $z=0.0$.} Consequently, the ratios {of $\Delta_{\rm min,DM}/\Delta_{\rm min, galaxies}$} in Figure~\ref{fig:integrateddeltaminmax}d) increase over time.


Figure~\ref{fig:vivintegrateddeltaminmax}a) shows the average integrated density contrasts within maximum-density spheres of ``voids-in-voids.'' Here, we impose a maximum radius of {$r\sim2.3R_{\rm eff}$} because some of the integrated ``void-in-void'' profiles have ridges that are underdense in both galaxies and dark matter compared to the average of the simulation. This means their integrated profiles continue to increase towards unity at radii much larger than the void radii. From Figure~\ref{fig:vivintegrateddeltaminmax}a), the value of $\Delta_{\rm max}$ for the dark matter shows a trend that is different to the trend in Figure~\ref{fig:integrateddeltaminmax}a). {Here, $\Delta_{\rm max}$ decreases over time from a maximum of $-0.024\pm0.009$ at $z=3.0$ to $-0.109\pm0.013$ at $z=0.0$.} The densities of the maximum-density spheres for galaxies {again fluctuate from redshift to redshift. These spheres have density contrasts that start at $-0.082\pm0.030$ at $z=3.0$ but change to $-0.063\pm0.029$ at $z=2.0$, $-0.113\pm0.017$ at $z=0.7$, $-0.082\pm0.020$ at $z=0.3$, and $-0.111\pm0.014$ at $z=0.0$}. The ratios {of $\Delta_{\rm min,DM}/\Delta_{\rm min, galaxies}$} in Figure~\ref{fig:vivintegrateddeltaminmax}b) thus increase over time. {Figure~\ref{fig:Reffmax}c) shows $R_{\rm max}$ for these spheres. With some exceptions, $R_{\rm max}$ for both profiles generally increase from $1.42\pm0.08R_{\rm eff}$ at $z=3.0$ to a maximum of $1.92\pm0.08R_{\rm eff}$ at $z=0.0$. Exceptions include the jump in $R_{\rm max}$ at $z=0.7$ and the $z=0.5$ and $z=0.2$ snapshots where $R_{\rm max}$ for the dark matter and galaxy profiles are not aligned. Thus, even ``voids-in-voids'' exhibit overdense ridges surrounding them, but as we will see for the ridges of ``voids-in-clouds,'' ``void-in-void'' ridges are far less pronounced and peak at higher values of $R_{\rm max}$ at redshifts $z\leq0.3$.}

Figure~\ref{fig:vivintegrateddeltaminmax}c) shows $\Delta_{\rm min}$ for ``voids-in-voids.''  Like the entire void population, these minimum-density spheres all have radii $r\leq0.25R_{\rm eff}$. From Figure~\ref{fig:vivintegrateddeltaminmax}c), $\Delta_{\rm min}$ values for galaxies remain constant at $\sim-0.98$ across time. {However, the value of $\Delta_{\rm min}$ for dark matter decreases from $-0.416\pm0.016$ at $z=3.0$ to a value of $-0.823\pm0.008$ at $z=0.0$.} Similar to what is seen in the entire void population, the ratios {of $\Delta_{\rm min,DM}/\Delta_{\rm min, galaxies}$} in Figure~\ref{fig:vivintegrateddeltaminmax}d) rise from $z=3.0$ to $z=0.0$.

Figure~\ref{fig:vicintegrateddeltaminmax}a) shows the average integrated number density contrasts within the maximum-density spheres of ``voids-in-clouds.'' Here, the maximum-density spheres have radii $r\leq1.5R_{\rm eff}$. The trends in $\Delta_{\rm max}$ for these voids closely resemble those of the entire void population in Figure~\ref{fig:integrateddeltaminmax}a). {Here, $\Delta_{\rm max}$ for the dark matter increases from an integrated density contrast of $0.064\pm0.008$ at $z=3.0$ to $0.281\pm0.035$ at $z=0.1$ and $0.2391\pm0.035$ at $z=0.0$. The average $\Delta_{\rm max}$ for galaxies shows a similar trend, changing from $0.196\pm0.026$ at $z=3.0$ to $0.334\pm0.042$ at $z=0.2$ to $0.267\pm0.039$ at $z=0.0$.} Thus, the ratios {of $\Delta_{\rm max,DM}/\Delta_{\rm max, galaxies}$} in Figure~\ref{fig:vicintegrateddeltaminmax}d) increase over time. {Figure~\ref{fig:Reffmax}d) shows the values of $R_{\rm max}$ for these maximum-density spheres. With the exception of $z=1.5$, $R_{\rm max}$ for the dark matter and density profiles align with each other. While the values of $R_{\rm max}$ fluctuate from snapshot to snapshot, the decrease in $\Delta_{\rm max}$ between redshifts $0.2-0.0$ again correlates with an increase in $R_{\rm max}$, again suggesting that the ridges of the voids tend towards larger radii at later times.}

Finally, Figure~\ref{fig:vicintegrateddeltaminmax}c) shows the evolution of $\Delta_{\rm min}$ for ``voids-in-clouds.'' As with the entire void population, these minimum-density spheres all have radii $r\leq0.25R_{\rm eff}$. ``Voids-in-clouds'' show the same trends as those seen in Figure~\ref{fig:integrateddeltaminmax}c) and Figure~\ref{fig:vivintegrateddeltaminmax}c). That is, $\Delta_{\rm min}$ for the dark matter decreases from {$-0.384\pm0.015$} at $z=3.0$ to a plateau of $-0.792\pm0.011$ at lower redshifts and the $\Delta_{\rm min}$ value for the galaxies remains at $-0.98$ over time. Thus, the ratios \text{of $\Delta_{\rm min,DM}/\Delta_{\rm min, galaxies}$} in Figure~\ref{fig:vicintegrateddeltaminmax}d) increase from $z=3.0$ to $z = 0.0$.

\section{Summary \& Discussion }
\label{sec:discussion}


We used the ZOBOV algorithm to identify voids in the galaxy fields of eleven snapshots of the \texttt{TNG300} simulation.  The snapshots span the redshift range $0 \le z \le 3$, corresponding to 11.65~Gyr of cosmic time.
We find that the galaxy distribution does not directly trace the dark matter distribution within the voids at any redshift. This fact is more evident at larger redshifts, where the centers of the voids are $\sim40\%$ more underdense in galaxies than dark matter and the ridges of the voids are $\sim15\%$ more overdense in galaxies than dark matter. We also find that, at all redshifts, there is a linear relationship between the galaxy and dark matter distributions that persists to radii $\sim 1.2R_{\rm eff}$. The slope of this relationship varies over time in accordance with the estimates of the linear clustering bias that exists within the simulation. 

The differences between the dark and luminous void density profiles decrease with decreasing redshift as dark matter leaves the centers of voids and populates the void ridges. {The density of galaxies in the centers of voids remains relatively constant over time while the density of galaxies around void ridges fluctuate between $\sim0.16$ at $z=3.0$ to $\sim0.29$ at $z=0.2$ and $\sim0.18$ at $z=0.0$.} That is, the centers of voids remain entirely devoid of galaxies, implying little to no net flow of galaxies into or out of \texttt{TNG300} voids between $z=3.0$ and $z=0.0$. {Our results for \texttt{TNG300} voids thus demonstrate that, as traced by galaxies from $z=3.0$ to $z=0.0$, the radial density profiles of voids experience only slight trends in their evolution within fluctuations.} 

We also identified distinct ``void-in-void'' and ``void-in-cloud'' populations in each snapshot. From this, we find the interiors of all voids, regardless of hierarchy, are nearly devoid of galaxies, and they become more underdense in dark matter over time. Differences that exist between the integrated profiles of ``voids-in-voids'' and ``voids-in-clouds'' occur primarily at the void ridges. For instance, the number density contrast of galaxies remains constant over time within the ridges of ``voids-in-clouds,'' while the number density contrast of dark matter particles increases over time and becomes equal to that of the galaxies at $z=0.0$. Between redshifts $z=3.0$ and $z=0.7$, the ridges of ``voids-in-voids'' tend to be more overdense in dark matter than in galaxies, while at lower redshifts, the ratios of galaxies to dark matter within the ridges of ``voids-in-voids'' approach unity. 

{The fluctuations in $\Delta_{\rm max}$ along void ridges could be indicative of a shift of the maximum density to slightly larger void radii at later times. We report such a shift in Figure~\ref{fig:Reffmax}b)-d), but it is more pronounced along the ridges of ``voids-in-clouds'' than it is for ``voids-in-voids''. For instance, when both galaxies and dark matter are used as tracers of ``voids-in-clouds,'' Figure~\ref{fig:Reffmax}d) shows that $R_{\rm max}$ increases from $1.42\pm0.08R_{\rm eff}$ at $z=0.2$ to $1.75\pm 0.08R_{\rm eff}$ at $z=0.0$. Meanwhile, for ``voids-in-voids,''  $R_{\rm max, DM}$ fluctuates between $1.58\pm0.08R_{\rm eff}$ at $z=0.3$, $1.92\pm0.08R_{\rm eff}$ at $z=0.2$, $1.75\pm0.08R_{\rm eff}$ at $z=0.1$, and $1.92\pm0.08R_{\rm eff}$ at $z=0.0$ while $R_{\rm max, galaxies}$ increases from $1.58\pm0.08R_{\rm eff}$ at $z=0.3$ to $1.75\pm0.08R_{\rm eff}$ at $z=0.2$ and $1.92\pm0.08R_{\rm eff}$ at $z=0.0$. Investigating this result with a larger sample of resolved voids is beyond the scope of this manuscript, but it would help clarify whether or not this is a result of selection effects and small number statistics.} 

At early to intermediate redshifts, ``voids-in-clouds'' experience an influx of dark matter into their ridges. In contrast, ``voids-in-voids'' experience a loss of dark matter from their ridges. This implies that care should be taken when attempting to use galaxies as a tracer of dark matter in observations of voids at intermediate to high redshifts, since the actual density contrast of dark matter in the ridges of voids may be either higher or lower than the density contrast of the galaxies depending on the void environment.


Most studies of voids in cosmological simulations limit their analyses to redshifts $z\lesssim0.7$ (see, e.g., \citealt{paillas2017}; \citealt{habouzit}; \citealt{davila2023}; \citealt{schuster2023}), so it is difficult to directly compare all our results to these studies. Despite the use of different voidfinder algorithms or trimming methods, however, our \texttt{TNG300} ZOBOV voids are comparable to voids found in the \texttt{EAGLE}, \texttt{HorizonAGN}, and \texttt{Magneticum} simulations, and it is reasonable to discuss our results in the context of some of these investigations. First, while \cite{pollina2017} focused on how $b_{\rm slope}$ varied as a function of void size in the $z=0.14$ snapshot of the  {\texttt{midres} run of the} \texttt{Magneticum Pathfinder} simulation, we studied the tracer-matter relationship from $z=3.0$ to $z=0.0$, finding a linear trend between $\delta_{\rm gal}$ and $\delta_{\rm DM}$ at all redshifts, and a slope that increases over time. {\cite{pollina2017} identified 36,430 voids in the galaxy field of the simulation, which covers a box that is $2,688h^{-1}$Mpc on a side.} Compared to the smallest voids that \cite{pollina2017} examined (i.e., those with radii between $20-30h^{-1}$Mpc), we report smaller values of $b_{\rm slope}$ at $z=0.1$ ($1.26\pm0.02$ in our study vs. $2.164\pm0.061$ in \citealt{pollina2017}). {This is likely a direct result of the sparser tracer population ($n_t\sim5\times10^{-4}h^3\rm{Mpc^{-3}}$ in their study vs. $n_t\sim3.9\times10^{-2}h^3\rm{Mpc^{-3}}$ in our study at $z=0.1$), but it could also be caused by the different resolution elements for the dark matter ($4.0\times10^{7}h^{-1}M_\odot$ for \texttt{TNG300} and $1.3\times10^{10}h^{-1}M_\odot$ for \texttt{Magneticum midres}) and baryons ($7.5\times10^{6}h^{-1}M_\odot$ for \texttt{TNG300} and $2.6\times10^{9}h^{-1}M_\odot$ for \texttt{Magneticum midres}), both of which can affect clustering bias (see, e.g., \citealt{nadathur2015c}; \citealt{pollina2016}}). However, our results agree with the general conclusions of \cite{pollina2017} in that, within voids, there is a linear relationship between $\delta_{\rm gal}$ and $\delta_{\rm DM}$, and the slope of this relationship is comparable to the clustering bias between the underlying matter distribution and a similarly selected tracer population.

\cite{habouzit} identified 36 voids with a median radius of $3.8h^{-1}${Mpc} and a median volume of $230$ $h^{-3}$Mpc$^3$ within {the galaxy field of} \texttt{HorizonAGN}. This simulation {has dark matter and stellar mass resolution elements of $8\times10^7M_\odot$ and $2\times10^6M_\odot$, respectively.} {Although \cite{habouzit} use a similar galaxy selection as we do (i.e., galaxies more massive than $2\times 10^8M_\odot$), the simulation} has a comoving box length of $100h^{-1}$Mpc, which is half the box length of \texttt{TNG300} (i.e., one eighth of the volume, making identification of the largest structures in \texttt{TNG300} not possible in \texttt{HorizonAGN}). \cite{habouzit} also used a version of ZOBOV that adopted somewhat different conditions for merging underdense regions together to build the overall void hierarchy,
and it is unclear how this will affect definitions of ``voids-in-voids'' and ``voids-in-clouds''. Thus, although our \texttt{TNG300} voids at $z=0.0$ are larger and more abundant than those in \texttt{HorizonAGN}, this is likely attributable to the larger volume of \texttt{TNG300} and differences in how underdense regions were merged together. 

\cite{paillas2017} used a 3DSVF to identify $\sim10^3$ voids in the {subhalo field of the} \texttt{EAGLE} simulation {at $z=0.0$}. {This simulation has a comoving box size of $100 \rm{Mpc^3}$ and has dark matter and baryon resolution elements of $9.7\times10^6M_\odot$ and $1.81\times10^6M_\odot$, respectively. These authors used a modified version of the 3DSVF that was developed by \cite{padilla2005}, where they modified the algorithm to test how the degree of allowable overlap between void regions affects their void catalogs. \cite{paillas2017} created void catalogs for a variety of tracer populations. Among these, the tracer population most comparable to our study consists of all subhalos with stellar masses $\geq10^8M_\odot$, containing 40,076 total subhalos and resulting in a tracer number density of $n_t\sim0.13h^{3}\rm{Mpc^{-3}}$ ($n_t\sim0.039h^{3}\rm{Mpc^{-3}}$ at $z=0.0$ in our study) and an average void size of $7.1$Mpc.} 
The most significant difference between their radial density profiles and those that we obtained is that the centers of \texttt{EAGLE} voids reach a minimum galaxy density contrast of $\sim-0.95$, a $5\%$ increase over what we find for the centers of \texttt{TNG300} voids. However, all other trends are consistent between the two simulations. For example, at $z=0.0$, the ridges of \texttt{EAGLE} voids and \texttt{TNG300}  voids reach similar galaxy number density contrasts. 
Similarly, both populations have interior dark matter density contrasts of $\sim-0.8$ and ridges that peak at $\sim0.15$. 

\cite{schuster2023} used ZOBOV to study dark matter radial density profiles of voids {identified in the halo and mass fields of the \texttt{midres}, \texttt{highres}, and \texttt{ultra-hr} runs of the \texttt{Magneticum} simulation. These runs have box lengths of $2,688h^{-1}$Mpc, $640h^{-1}$Mpc, and $48h^{-1}$Mpc, dark matter particle counts of $2\times4,536^3$, $2\times2,880^3$, and $2\times576^3$, and dark matter (baryon) resolution elements of $1.3\times10^{10}h^{-1}M_\odot$ ($2.6\times10^9h^{-1}M_\odot$), $6.9\times10^8h^{-1}M_\odot$ ($1.4\times10^8h^{-1}M_\odot$), and $3.6\times10^7h^{-1}M_\odot$ ($7.3\times10^6h^{-1}M_\odot$), respectively. When halos are used as tracers, \cite{schuster2023} apply mass cuts of $1.0\times10^{12}h^{-1}M_\odot$, $1.0\times10^{11}h^{-1}M_\odot$, and $1.3\times10^9h^{-1}M_\odot$, giving tracer number densities of $3.2\times10^{-3}h^3\rm{Mpc^{-3}}$, $3.1\times10^{-2}h^3\rm{Mpc^{-3}}$, and $1.2h^3\rm{Mpc^{-3}}$, respectively ($n_t\sim4.0\times10^{-2}h^3\rm{Mpc^{-3}}$ in our study at $z=0.0$). Although their \textit{merged} void sample was trimmed with an imposed maximum zone-linking density, the voids they analyzed are generally similar to the ones we identify.} They found that the dark matter distribution within voids varied as a function of void size and shape. Compared to our dark matter profile at $z=0.0$, their stacked profile is similar, despite resolution differences between \texttt{TNG300} and \texttt{Magneticum}. Similarly, \cite{davila2023} used a 3DSVF to examine integrated ``void-in-void'' and ``void-in-cloud'' profiles of \texttt{TNG300} voids. At $z=0.0$, the interiors and ridges of our ``void-in-void'' and ``void-in-cloud'' profiles are nearly identical to theirs, reaching similar minima and maxima despite different choices for the implementation of the voidfinding algorithm.

Using observational data, \cite{douglass2023} identified voids in the Sloan Digital Sky Survey (SDSS; \citealt{SDSS}) DR7 \citep{abazajian2009} and investigated how the profiles of voids identified with 3DSVFs differed from those identified with ZOBOV. In particular, the 3DSVF VoidFinder (\citealt{elad1997}; \citealt{hoyle2004}) identified voids with flatter interior profiles and ridges that were only slightly denser than the surrounding environment. Despite using the other two commonly used methods for trimming raw ZOBOV catalogs, their $z=0.0$ ZOBOV density profiles closely resemble the $z=0.0$ luminous density profiles that we present here.

\cite{hamaus2020} used ZOBOV to study void-galaxy cross-correlation functions in the SDSS Baryon Oscillation Spectroscopic Survey (BOSS; \citealt{BOSS}), where they imposed a minimum void radius to trim their raw catalogs. Compared to the voids that we identified at redshift $z\sim0.5$, the ridges of their monopole profiles are similar, but their interiors are slightly more overdense in galaxies ($\xi_0\sim-0.90$). \cite{nadathur2019} and \cite{woodfinden2022} also obtained cross-correlation functions for SDSS BOSS and SDSS extended BOSS (eBOSS; \citealt{eBOSS}) Data Releases. Both studies identified voids with ZOBOV, trimming their catalogs by imposing a minimum void radius. They found voids out to $z=0.7$ with similar interiors to those that we present but with ridges that only reach values of $\xi_0\sim0.05$.

Forthcoming surveys will map the galaxy distribution in the observed universe to unprecedented depths, encompassing enormous volumes of space and dramatically enhancing our understanding of the growth of structure over the past $\sim 12$~Gyr, including the evolution of voids over cosmic time.  Our results for \texttt{TNG300} voids demonstrate that, as traced by galaxies, the radial density profiles of voids {fluctuates} from $z=3.0$ to $z=0.0$, and the relationship between void galaxy density contrast and dark matter density contrast is consistent with linear bias over this time period.  If $\Lambda$CDM is the correct model of structure formation, then, future void surveys should yield similar conclusions for voids in the observed universe.  

While using galaxies to obtain radial density profiles for voids as a function of redshift should be relatively straightforward (given sufficiently deep observations), testing the evolution of the void galaxy-dark matter bias relationship that we obtained will depend on the ability to directly constrain the mass density profiles of voids over a similar redshift range. This may be possible in future because,   although highly challenging to carry out, 
(see, e.g., \citealt{izumi2013}; \citealt{barreira2015}; \citealt{sanchez2016cosmic}; \citealt{paillas2019}; \citealt{bonici2023}), 
precise weak lensing studies of voids hold the promise to constrain the relationship between void galaxies and dark matter in much the same way that weak lensing studies have constrained the galaxy-matter cross-correlation function for galaxies (see \citealt{more2023} and references therein).

\section*{Acknowledgements}
{We are grateful to the anonymous reviewer for helpful comments and suggestions that improved the manuscript.} This work was partially supported by the National Science Foundation grant AST-2009397. The \texttt{TNG} simulations were undertaken with compute time awarded by the Gauss Centre for Supercomputing (GCS) under GCS Large-Scale Projects GCS-ILLU and GCS-DWAR on the GCS share of the supercomputer Hazel Hen at the High Performance Computing Center Stuttgart (HLRS), as well as on the machines of the Max Planck Computing and Data Facility (MPCDF) in Garching, Germany. In addition, we are pleased to acknowledge that the computational work reported on in this paper was performed on the Shared Computing Cluster, administered by Boston University’s Research Computing Services, and located at the Massachusetts Green High Performance Computing Center in Holyoke, MA. OC acknowledges support from the Penn State Extraterrestrial Intelligence Center. BM acknowledges support from Northeastern University's Future Faculty Postdoctoral Fellowship program. 



\bibliography{bibliography}

\label{lastpage}
\end{document}